\documentclass[12pt, reqno]{amsart}

\usepackage[margin=.9in, a4paper, foot=.3in]{geometry}

\let\oldtocsection=\tocsection

\let\oldtocsubsection=\tocsubsection

\let\oldtocsubsubsection=\tocsubsubsection

\renewcommand{\tocsection}[2]{\hspace{0em}\oldtocsection{#1}{#2}}
\renewcommand{\tocsubsection}[2]{\hspace{1em}\oldtocsubsection{#1}{#2}}
\renewcommand{\tocsubsubsection}[2]{\hspace{2em}\oldtocsubsubsection{#1}{#2}}

\usepackage{cite}

\usepackage{graphicx, color}

\definecolor{darkblue}{rgb}{0,0,.5}
\definecolor{darkred}{rgb}{.5,0,0}
\definecolor{darkgreen}{rgb}{0,0.5,0}

\usepackage{hyperref}
\hypersetup {
    pdfstartview=FitH,
    colorlinks,
    citecolor=darkgreen,
    linkcolor=darkred,
    urlcolor=darkblue
}

\usepackage[noBBpl,slantedGreek]{mathpazo}
\usepackage[mathcal]{euscript}

\usepackage{amssymb}

\makeatletter
\def\section{\@startsection{section}{1}%
  \z@{-.7\linespacing\@plus -\linespacing}{.5\linespacing}%
  {\normalfont\scshape\centering}}
\def\subsection{\@startsection{subsection}{2}%
  \z@{-.5\linespacing\@plus -.7\linespacing}{.5em}%
  {\normalfont\bfseries\mathversion{bold}}}
\makeatother

\allowdisplaybreaks

\numberwithin{equation}{section}

\newcommand {\id}{\mathrm{id}}
\newcommand {\Osc}{\mathrm{Osc}}
\newcommand {\rme}{\mathrm e}

\newcommand {\bbC}{\mathbb C}
\newcommand {\bbD}{\mathbb D}
\newcommand {\bbL}{\mathbb L}
\newcommand {\bbM}{\mathbb M}
\newcommand {\bbN}{\mathbb N}
\newcommand {\bbO}{\mathbb O}

\newcommand {\bbQ}{\mathbb Q}
\newcommand {\bbS}{\mathbb S}
\newcommand {\bbT}{\mathbb T}

\newcommand {\bbZ}{\mathbb Z}

\newcommand {\calA}{\mathcal A}
\newcommand {\calB}{\mathcal B}
\newcommand {\calC}{\mathcal C}
\newcommand {\calD}{\mathcal D}

\newcommand {\calK}{\mathcal K}
\newcommand {\calL}{\mathcal L}
\newcommand {\calM}{\mathcal M}

\newcommand {\calQ}{\mathcal Q}
\newcommand {\calR}{\mathcal R}
\newcommand {\calT}{\mathcal T}

\newcommand {\hcalL}{\widehat{\mathcal L}}
\newcommand {\hgothh}{\widehat{\mathfrak h}}
\newcommand {\hlslii}{\widehat{\mathcal L}(\mathfrak{sl}_2)}

\newcommand {\hbbQ}{\widehat{\mathbb Q}{}}
\newcommand {\hbbT}{\widehat{\mathbb T}{}}

\newcommand {\tcalL}{\widetilde{\mathcal L}}
\newcommand {\tgothh}{\widetilde{\mathfrak h}}

\newcommand {\gothh}{\mathfrak h}
\newcommand {\gothg}{\mathfrak g}
\newcommand {\gothgl}{\mathfrak{gl}}
\newcommand {\gothglii}{\mathfrak{gl}_2}

\newcommand {\gothslii}{\mathfrak{sl}_2}

\newcommand {\op}{\mathrm{op}}

\newcommand {\ssplus}{{\scriptscriptstyle +}}
\newcommand {\ssminus}{{\scriptscriptstyle -}}
\newcommand {\tlslii}{\widetilde{\mathcal L}(\mathfrak{sl}_2)}

\newcommand {\uqbp}{\mathrm U_q(\mathfrak b_+)}
\newcommand {\uqbm}{\mathrm U_q(\mathfrak b_-)}
\newcommand {\uqhlslii}{\mathrm U_q(\widehat{\mathcal L}(\mathfrak{sl}_2))}

\newcommand {\uqglii}{\mathrm U_q(\mathfrak{gl}_2)}

\newcommand {\uqslii}{\mathrm U_q(\mathfrak{sl}_2)}

\newcommand {\uqlslii}{\mathrm U_q(\mathcal L(\mathfrak{sl}_2))}
\newcommand {\uqlsliii}{\mathrm U_q(\mathcal L(\mathfrak{sl}_3))}
\newcommand {\uqtlslii}{\mathrm U_q(\widetilde{\mathcal L}(\mathfrak{sl}_2))}

\DeclareMathOperator {\End}{End}

\DeclareMathOperator {\tr}{tr}

\title{Quantum groups and functional relations for lower rank}

\author[Kh. S. Nirov]{\vskip .2em Khazret S. Nirov}
\address{Institute for Nuclear Research of the Russian Academy of Sciences, 60th October Ave 7a, 117312 Moscow, Russia}
\curraddr{Fachbereich C -- Physik, Bergische Universit\"at Wuppertal, 42097 Wuppertal, Germany}
\email{nirov@uni-wuppertal.de}

\author[A. V. Razumov]{Alexander V. Razumov}
\address{Institute for High Energy Physics, 142281 Protvino, Moscow region, Russia}
\email{Alexander.Razumov@ihep.ru}

\begin{document}

\addtolength {\jot}{3pt}

\begin{abstract}
A detailed construction of the universal integrability objects related to the integrable systems associated with the quantum group $\uqlslii$ is given. The full proof of the functional relations in the form independent of the representation of the quantum group on the quantum space is presented. The case of the general gradation and general twisting is treated. The specialization of the universal functional relations to the case when the quantum space is the state space of a discrete spin chain is described. This is a degression of the corresponding consideration for the case of the quantum group $\uqlsliii$ with an extensions to the higher spin case.
\end{abstract}

\maketitle

\tableofcontents

\section{Introduction}

The modern approach to the study of quantum integrable systems is based on the concept of the transfer matrix, or transfer operator, and the main problem here is to find its eigenvalues. The most productive method to do this is the Bethe ansatz~\cite{Bet31}. Unfortunately, it does not work for all cases which considered as integrable. More general method was invented by Baxter, see, for example \cite{Bax82, Bax04}. He proposed to consider, together with the transfer operator, an auxiliary operator, called the $Q$-operator. The transfer operator and $Q$-operator satisfy some difference equation called the Baxter's functional $TQ$-relation. At present, the transfer operators and $Q$-operators are constructed as traces of the corresponding monodromy operators and $L$-operators. We call all the operators, mentioned above, the integrability objects.

It was noted by Bazhanov, Lukyanov and Zamolodchikov \cite{BazLukZam96, BazLukZam97, BazLukZam99} that the integrability objects can be constructed starting with the universal $R$-matrix of the quantum group related to the quantum integrable system under consideration. Here the corresponding functional relations follow from the properties of representations of the quantum group. The method was used for construction of $R$-operators \cite{KhoTol92, LevSoiStu93, ZhaGou94, BraGouZhaDel94, BraGouZha95, BooGoeKluNirRaz10, BooGoeKluNirRaz11}, monodromy operators and $L$-operators \cite{BazLukZam96, BazLukZam97, BazLukZam99, BazTsu08, BooGoeKluNirRaz10, BooGoeKluNirRaz11, BooGoeKluNirRaz13, Raz13, BooGoeKluNirRaz14a, BooGoeKluNirRaz14b}, and for the proof of functional relations \cite{BazLukZam99, BazHibKho02, Koj08, BazTsu08, BooGoeKluNirRaz14a, BooGoeKluNirRaz14b}.

The notion of quantum group was introduced by Drinfeld and Jimbo \cite{Dri85, Dri87, Jim85}. Roughly speaking, it is a special kind of Hopf algebras. The universal $R$-matrix is an element of the tensor product of two copies of the quantum group, see, for example \cite{ChaPre91}. The integrability objects are determined by a choice of representations for the factors of that tensor product. The choice made for the first factor determines an integrability object, and for the second factor a concrete integrable model. It is common to call the representation space of the first factor an auxiliary space, and the representation space of the second one the quantum space. However, the roles of the factors can be interchanged. It appeared productive to fix representation only for the first factor, see, for example \cite{AntFei97, BazTsu08, BooGoeKluNirRaz13, BooGoeKluNirRaz14a,BooGoeKluNirRaz14b}. We call the arising integrability objects and functional relations universal ones.

In this paper we consider quantum integrable systems related to the quantum group $\uqlslii$. For this case the main source of the representations of the quantum group is the Jimbo's homomorphism \cite{Jim86a}. In the paper \cite{BooGoeKluNirRaz14a} the Jimbo's homomorphism was defined as a mapping from $\uqlslii$ to $\uqslii$. It appears that formulas become simpler and the algebraic basis more transparent if one considers the Jimbo's homomorphism as a mapping from $\uqlslii$ to $\uqglii$. Therefore, we rederive the universal functional relations using another form of the Jimbo's homomorphism. Here we use a universal approach to $TQ$- and $TT$-relations proposed in the paper \cite{BooGoeKluNirRaz14b}. We find explicit expressions for the monodromy operators and $L$-operators for the spin chains of `spin' $1/2$ particles. Then, using the fusion procedure, we construct the expressions for the `spin' 1 case. The results of the fusion procedure allow us to define integrability objects which are Laurent polynomials on some power of the spectral parameter. Finally, we specialize the functional relations to the case of spin chains defined by a choice of arbitrary finite dimensional representation for the quantum space and write them in terms of polynomial objects.

Depending on the sense of the deformation parameter $q$, there are at least three definitions of a quantum group. According to the first definition, $q = \exp \hbar$, where $\hbar$ is an indeterminate, according to the second one, $q$ is indeterminate, and according to the third one, $q = \exp \hbar$, where $\hbar$ is a complex number. In the first case the quantum group is a $\bbC[[\hbar]]$-algebra, in the second case a $\bbC(q)$-algebra, and in the third case it is just a complex algebra. To construct integrability objects one uses trace operations on the quantum group under consideration. To define traces it seems convenient to use the third definition of the quantum group. Therefore, we define the quantum group as a $\bbC$-algebra, see, for example, the books \cite{JimMiw95, EtiFreKir98}.

We denote by $\calL(\gothg)$ the loop Lie algebra of a finite dimensional simple Lie algebra $\gothg$, by $\tcalL(\gothg)$ its standard central extension, and by $\hcalL(\gothg)$ the Lie algebra $\tcalL(\gothg)$ endowed with a special differentiation, see, for example, the book by Kac \cite{Kac90}. The symbol~$\bbN$ means the set of natural numbers and the symbol $\bbZ_+$ the set of non-negative integers.

Depending on the context, the symbol `1' means the integer one, the unit of an algebra, or the unit matrix. The symbol $\otimes$ denotes the tensor product of vector spaces and algebras and the Kronecker product of matrices with commuting or noncommuting entries.

Below we use the notation
\begin{equation*}
\kappa_q = q - q^{-1},
\end{equation*}
so that the definition of the $q$-deformed number can be written as 
\begin{equation*}
[\nu]_q = \frac{q^\nu - q^{- \nu}}{q - q^{-1}} = \kappa_q^{-1}(q^\nu - q^{-\nu}), \qquad \nu \in \bbC.
\end{equation*}
To construct integrability objects one uses spectral parameters. They are introduced by defining a $\bbZ$-gradation of the quantum group under consideration. We consider integrable systems related to the quantum group $\uqlslii$. In this case a $\bbZ$-gradation is determined by two integers $s_0$, $s_1$. We often use the notation $s = s_0 + s_1$.

\section{Integrability objects}

\subsection{\texorpdfstring{Quantum group $\uqglii$}{Quantum group Uq(gl2)}} \label{ss:qguqslii}

To construct integrability objects one uses appropriate homomorphisms of the quantum group under consideration. For the case of the quantum group $\uqlslii$ the most important homomorphism is the Jimbo's homomorphism from $\uqlslii$ to the quantum group $\uqglii$. Therefore, we first remind the definition of $\uqglii$ and discuss its representations, and then proceed to $\uqlslii$. 

\subsubsection{Definition}

Denote by $\gothg$ the standard Cartan subalgebra of the Lie algebra $\gothgl_2$ and by $G_i = E_{ii}$, $i = 1, 2$, the elements forming the standard basis of $\gothg$.\footnote{We use the usual notation $E_{ij}$ for the matrix units.} The root system of $\gothgl_2$ relative to $\gothg$ consists of two roots $\alpha$ and $-\alpha$ and we have
\begin{equation}
\alpha(G_1) = 1, \qquad \alpha(G_2) = -1. \label{alphah}
\end{equation}

The Lie algebra $\gothslii$ is a subalgebra of $\gothglii$, and the standard Cartan subalgebra $\gothh$ of $\gothslii$ is a subalgebra of $\gothg$. Here the standard Cartan generator $H$ of $\gothslii$ is
\begin{equation}
H = G_1 - G_2, \label{hk}
\end{equation}
and we have
\begin{equation*}
\alpha(H) = 2.
\end{equation*}

We define the quantum group $\uqglii$ as a unital associative $\bbC$-algebra generated by the elements $E$, $F$ and $q^X$, $X \in \gothg$, with the relations
\begin{gather}
q^0 = 1, \qquad q^{X_1} q^{X_2} = q^{X_1 + X_2}, \label{xx} \\
q^X E \, q^{-X} = q^{\alpha(X)} E, \qquad q^X F \, q^{-X} = q^{- \alpha(X)} F, \label{xexf} \\
[E, F] = \kappa_q^{-1} \, (q^H - q^{-H}). \label{ef}
\end{gather}
Note that $q^X$ is just a convenient notation. There are no elements of $\uqglii$ corresponding to the elements of $\gothg$. In fact, this notation means a set of elements of $\uqglii$ parametrized by $\gothg$. It is convenient to use the notations
\begin{equation*}
q^{X + \nu} = q^\nu q^X
\end{equation*}
and
\begin{equation}
[X + \nu]_q = \kappa_q^{-1} \, (q^{X + \nu}- q^{ -X -\nu}) = \kappa_q^{-1} \, (q^{\nu} q^X - q^{-\nu} q^{-X}) \label{xnq}
\end{equation}
for any $X \in \gothg$ and $\nu \in \bbC$. Here equation (\ref{ef}) takes the form
\begin{equation*}
[E, F] = [H]_q.
\end{equation*}
Similar notations are used below for the case of the quantum groups $\uqtlslii$ and $\uqlslii$.

With respect to the properly defined coproduct, counit and antipode the quantum group $\uqglii$ is a Hopf algebra.

The quantum group $\uqslii$ is a Hopf subalgebra of $\uqglii$ generated by $E$, $F$ and $q^X$, $X \in \gothh$.

\subsubsection{Highest weight modules} \label{sss:hwm}

Let $\lambda$ be an element of $\gothg^*$. We identify $\lambda$ with the set of its components $(\lambda_1, \lambda_2)$ with respect to the dual basis of the basis $\{G_i\}$. In fact we have
\begin{equation*}
\lambda_1 = \lambda(G_1), \qquad \lambda_2 = \lambda(G_2).
\end{equation*}
For the simple root $\alpha$ we obtain the identification
\begin{equation*}
\alpha = (1, \, -1).
\end{equation*}
The Verma $\uqglii$-module $\widetilde V^\lambda$  is the $\uqglii$-module with the highest weight vector $v_0$ satisfying the relations
\begin{equation}
q^{\nu G_1} v_0 = q^{\nu \lambda_1} v_0, \qquad q^{\nu G_2} v_0 = q^{\nu \lambda_2} v_0, \qquad E v_0 = 0. \label{hwr}
\end{equation}
It is convenient to denote
\begin{equation}
\mu = \lambda_1 - \lambda_2, \label{mu}
\end{equation}
so that for the generators $q^{\nu H}$ of $\uqslii$ we have
\begin{equation*}
q^{\nu H} v_0 = q^{\nu \mu} v_0.
\end{equation*}

The vectors
\begin{equation}
v_n = F^n v_0, \label{vnii}
\end{equation}
where $n \in \bbZ_+$, form a basis of $\widetilde V^\lambda$. Let us describe the action of the generators of $\uqglii$ on the elements of this basis.

Using (\ref{xexf}) and (\ref{alphah}), we obtain
\begin{equation*}
q^{\nu G_1} F = q^{- \nu} F q^{\nu G_1}, \qquad q^{\nu G_2} F = q^\nu F q^{\nu G_2}.
\end{equation*}
Now, taking into account (\ref{hwr}), we see that
\begin{equation*}
q^{\nu G_1} v_n = q^{\nu(\lambda_1 - n)} v_n, \qquad q^{\nu G_2} v_n = q^{\nu(\lambda_2 + n)} v_n.
\end{equation*}
Further, it follows directly from (\ref{vnii}) that
\begin{equation*}
F v_n = v_{n + 1}.
\end{equation*}
Finally, using (\ref{ef}), (\ref{xexf}) and (\ref{alphah}) we come to the equation
\begin{equation*}
E v_n = [n]_q [\lambda_1 - \lambda_2 - n + 1]_q v_{n - 1}.
\end{equation*}
which can be also written as
\begin{equation*}
E v_n = [n]_q [\mu - n + 1]_q v_{n - 1}.
\end{equation*}

We denote the representation of $\uqglii$ corresponding to the module $\widetilde V^\lambda$ by $\widetilde \pi^\lambda$. When the number $\mu$ defined by (\ref{mu}) is a non-negative integer, the infinite dimensional module $\widetilde V^{(\lambda_1, \lambda_2)}$ has an infinite dimensional submodule $\widetilde V^{(\lambda_2 - 1, \lambda_1 + 1)}$. The corresponding quotient module is $(\lambda_1 - \lambda_2 + 1)$-dimensional. We denote this finite dimensional $\uqglii$-module by $V^\lambda$ and the corresponding representation by $\pi^\lambda$.

For the quantum group $\uqglii$ there are two independent quantum Casimir elements which we choose in the form
\begin{align}
& C^{(1)} = q^{- 2 G_1 - 1} + q^{- 2 G_2 + 1} + \kappa_q^2 F E q^{- G_1 - G_2} \notag \\
& \hspace{15em} {} = q^{- 2 G_1 + 1} + q^{- 2 G_2 - 1} + \kappa_q^2 E F q^{- G_1 - G_2}, \label{qciia} \\
& C^{(2)} = q^{- 2 G_1 - 2 G_2}. \label{qciib}
\end{align}
For the representations $\widetilde \pi^\lambda$ and $\pi^\lambda$ one has
\begin{align}
& \widetilde \pi^\lambda(C^{(1)}) = \pi^\lambda(C^{(1)}) = q^{- 2 \lambda_1 - 1} + q^{- 2 \lambda_2 + 1}, \label{plc1} \\
& \widetilde \pi^\lambda(C^{(2)}) = \pi^\lambda(C^{(2)}) = q^{- 2 \lambda_1 - 2 \lambda_2}. \label{plc2}
\end{align}

\subsubsection{Appearance of BGG resolution}

As we noted above, when the number $\lambda_1 - \lambda_2$ is a non-negative integer, the infinite dimensional module $\widetilde V^{(\lambda_1, \lambda_2)}$ has an infinite dimensional submodule $\widetilde V^{(\lambda_2 - 1, \lambda_1 + 1)}$. The corresponding quotient module is $(\lambda_1 - \lambda_2 + 1)$-dimensional. In fact, we have an exact sequence
\begin{equation}
0 \longrightarrow \widetilde V^{(\lambda_2 - 1, \lambda_1 + 1)} \overset{i}{\longrightarrow} \widetilde V^{(\lambda_1, \lambda_2)} \overset{p}{\longrightarrow} V^{(\lambda_1, \lambda_2)} \longrightarrow 0, \label{fdm}
\end{equation}
where $i$ is the inclusion homomorphism and $p$ the canonical projection. Let us show that this is an example of quantum Bernstein--Gelfand--Gelfand (BGG) resolution \cite{Ros91}.

Recall first some definitions and properties of the necessary objects. We have denoted the standard Cartan subalgebra of the Lie algebra $\gothglii$ by $\gothg$. The Weyl group $W$ of the root system of $\gothglii$ is generated by the reflection $r: \gothg^* \to \gothg^*$ defined by the equation
\begin{equation*}
r(\lambda) = \lambda - \lambda(H) \alpha.
\end{equation*}
The minimal number of generators $r$ necessary to represent an element $w \in W$ is said to be the length of $w$ and is denoted by $\ell(w)$. It is assumed that the identity element has the length equal to $0$.

Let $\{\gamma_i\}$ be a dual basis of the standard basis $\{G_i\}$ of $\gothg$. Using (\ref{alphah}), it is easy to see that
\begin{equation*}
\alpha = \gamma_1 - \gamma_2.
\end{equation*}
One can get convinced that
\begin{equation*}
r (\gamma_1) = \gamma_2, \qquad r (\gamma_2) = \gamma_1.
\end{equation*}
Identifying an element of $\gothg^*$ with the set of its components with respect to the basis $\{\gamma_i\}$, we see that the reflection $r$ transposes the first and second components. It is clear that the whole Weyl group $W$ can be identified with the symmetric group $\mathrm S_2$. Here $(-1)^{\ell(w)}$ is evidently the sign of the permutation corresponding to the element $w \in W$.  The order of $W$ is equal to two. There are one element of length $0$ and one element of length $1$. Denote
\begin{equation*}
U_k = \bigoplus_{\substack{w \in W \\ \ell(w) = k}} \widetilde V^{w \cdot \lambda},
\end{equation*}
where $w \cdot \lambda$ means the affine action of $w$ defined as
\begin{equation*}
w \cdot \lambda = w(\lambda + \rho) - \rho
\end{equation*}
with $\rho = \alpha / 2$ the half-sum of positive roots. One can verify that
\begin{equation}
r \cdot (\lambda_1, \lambda_2) = (\lambda_2 - 1, \lambda_1 + 1). \label{rll}
\end{equation}
Hence, we have
\begin{equation*}
U_0 = \widetilde V^{(\lambda_1, \lambda_2)}, \qquad U_1 = \widetilde V^{(\lambda_2 - 1, \lambda_1 + 1)}. 
\end{equation*}
In the case under consideratin the quantum version of the Bernstein--Gelfand--Gelfand resolution is the following exact sequence of $\uqglii$-modules and $\uqglii$-ho\-mo\-mor\-phisms:
\begin{equation}
0 \longrightarrow U_1 \overset{\varphi_1}{\longrightarrow} U_0 \overset{\varphi_0}{\longrightarrow} U_{-1} \longrightarrow 0, \label{bgg}
\end{equation}
where $U_{-1} = V^\lambda$. Assuming that $\varphi_0 = p$ and $\varphi_1 = i$, we see that (\ref{fdm}) coincides with (\ref{bgg}). 

\subsection{\texorpdfstring{Quantum group $\uqlslii$}{Quantum group Uq(L(sl2))}}

\subsubsection{Definition}

We start with the quantum group $\uqhlslii$. Recall that the Cartan subalgebra of $\hlslii$ is
\begin{equation*}
\hgothh = \gothh \oplus \bbC c \oplus \bbC d,
\end{equation*}
where $\gothh = \bbC H$ is the standard Cartan subalgebra of $\gothslii$, $c$ the central element, and $d$ the differentiation \cite{Kac90}. Define the Cartan elements
\begin{equation*}
h_0 = c - H, \qquad h_1 = H,
\end{equation*}
so that one has
\begin{equation}
c = h_0 + h_1 \label{c}
\end{equation}
and
\begin{equation*}
\hgothh = \bbC h_0 \oplus \bbC h_1 \oplus \bbC d.
\end{equation*}
The simple roots $\alpha_i \in \hgothh^*$, $i = 0$, $1$, are given by the equation
\begin{equation*}
\alpha_j(h_i) = \widetilde a_{ij}, \qquad \alpha_0(d) = 1, \qquad \alpha_1(d) = 0,
\end{equation*}
where
\begin{equation*}
(\widetilde a_{ij}) = \left(\begin{array}{rr}
2 & -2 \\
-2 & 2
\end{array} \right)
\end{equation*}
is the Cartan matrix of the Lie algebra $\hlslii$.

As before, let $\hbar$ be a complex number and $q = \exp \hbar$. The quantum group $\uqhlslii$  is a unital associative $\bbC$-algebra generated by the elements $e_i$, $f_i$, $i = 0, 1$, and $q^x$, $x \in \hgothh$, with the relations
\begin{gather}
q^0 = 1, \qquad q^{x_1} q^{x_2} = q^{x_1 + x_2}, \label{lxx} \\
q^x e_i q^{-x} = q^{\alpha_i(x)} e_i, \qquad q^x f_i q^{-x} = q^{-\alpha_i(x)} f_i, \\
[e_i, f_j] = \delta_{ij} [h_i]_q
\end{gather}
satisfied for all $i$ and $j$, and the Serre relations
\begin{align}
& e_i^3 e_j^{\mathstrut} - [3]_q  e_i^2 e_j^{\mathstrut} e_i^{\mathstrut} + [3]_q  e_i{\mathstrut} e_j^{\mathstrut} e_i^2 - e_j^{\mathstrut} e_i^3 = 0, \\
& f_i^3 f_j^{\mathstrut} - [3]_q  f_i^2 f_j^{\mathstrut} f_i^{\mathstrut} + [3]_q  f_i{\mathstrut} f_j^{\mathstrut} f_i^2 - f_j^{\mathstrut} f_i^3 = 0 \label{lsr}
\end{align}
satisfied for all distinct $i$ and $j$.

The quantum group $\uqhlslii$ is a Hopf algebra with the comultiplication $\Delta$, the antipode $S$, and the counit $\varepsilon$ defined by the
relations
\begin{gather}
\Delta(q^x) = q^x \otimes q^x, \qquad \Delta(e_i) = e_i \otimes 1 + q^{- h_i} \otimes e_i, \qquad \Delta(f_i) = f_i \otimes q^{h_i} + 1 \otimes f_i, \label{cmul} \\
S(q^x) = q^{- x}, \qquad S(e_i) = - q^{h_i} e_i, \qquad S(f_i) = - f_i \, q^{- h_i}, \\
\varepsilon(q^x) = 1, \qquad \varepsilon(e_i) = 0, \qquad \varepsilon(f_i) = 0. \label{cu}
\end{gather}

The quantum group $\uqlslii$ can be defined as the quotient algebra of $\uqhlslii$ by the two-sided ideal generated by the elements of the form $q^{\nu c} - 1$ or of the form $q^{\nu d} - 1$ with $\nu \in \bbC^\times$. Introduce the notation
\begin{equation*}
\tgothh = \gothh \oplus \bbC c = \bbC h_0 \oplus \bbC h_1.
\end{equation*}
It is convenient to consider the quantum group $\uqlslii$ as a $\bbC$-algebra generated by the elements $e_i$, $f_i$, $i = 0, 1$, and $q^x$, $x \in \widetilde{\gothh}$, with relations (\ref{lxx})--(\ref{lsr}) and additional relation
\begin{equation}
q^{\nu c} = 1, \qquad \nu \in \bbC^\times. \label{qh0h1}
\end{equation}
It is a Hopf algebra with the Hopf structure defined by (\ref{cmul})--(\ref{cu}). One of the reasons to use the quantum group $\uqlslii$ instead of $\uqhlslii$ is that in the case of $\uqhlslii$ we have no interesting finite dimensional representations.

\subsubsection{Cartan--Weyl generators}

An element $a$ of $\uqlslii$ is called a root vector corresponding to a root $\gamma$ of $\tlslii$ if
\begin{equation*}
q^x a \, q^{-x} = q^{\gamma(x)} a
\end{equation*}
for all $x \in \tgothh$. It is clear that $e_i$ and $f_i$ are root vectors corresponding to the roots $\alpha_i$ and $- \alpha_i$. One can find linearly independent root vectors corresponding to all roots of $\tlslii$. These vectors, together with the elements $q^x$, $x \in \tgothh$ are called Cartan--Weyl generators of $\uqlslii$. It appears that the ordered monomials constructed from the Cartan--Weyl generators form a Poincar\'e--Birkhoff--Witt basis of $\uqlslii$. 

To construct root vectors we follow the papers \cite{TolKho92, KhoTol93, KhoStoTol95}. Here we denote the root vector corresponding to a positive root $\gamma$ by $e_\gamma$ and the root vector corresponding to a negative root $- \gamma$ by $f_\gamma$. The system of positive roots of $\tlslii$ is
\begin{equation*}
\widetilde \triangle_+ = \{ \alpha + k \delta \mid k \in \bbZ_+ \} \cup \{ k \delta \mid k \in \bbN \} \cup \{ (\delta - \alpha) + k \delta \mid k \in \bbZ_+ \},
\end{equation*}
where
\begin{equation*}
\delta = \alpha_0 + \alpha_1, \qquad \alpha = \alpha_1.
\end{equation*}
The full system of roots $\widetilde \triangle$ is the union of the systems of positive and negative roots, $\widetilde \triangle = \widetilde \triangle_+ \cup (- \widetilde \triangle_+)$.
We fix the following normal order of $\widetilde \triangle_+$:
\begin{equation*}
\alpha, \, \alpha + \delta, \, \ldots, \, \alpha + k \delta, \, \ldots, \, \delta, \, 2 \delta, \,
\ldots, \, k \delta, \, \ldots, \, \ldots, \, (\delta - \alpha) + k \delta, \, \ldots, \,
(\delta - \alpha) + \delta, \, \delta - \alpha,
\end{equation*}
see the paper \cite{TolKho92}.

We have root vectors corresponding to the roots $\alpha$ and $\delta - \alpha$,
\begin{equation*}
e_\alpha = e_1, \qquad e_{\delta - \alpha} = e_0.
\end{equation*}
Then we construct a root vector corresponding to the root $\delta$ by the relation
\begin{equation}
e'_\delta = e_\alpha \, e_{\delta - \alpha} - q^{-2} e_{\delta - \alpha} \, e_\alpha. \label{e2}
\end{equation}
The prime here means that below we redefine the vectors corresponding to the roots $k \delta$. The next step is to construct vectors corresponding to the roots $\alpha + k \delta$ and $(\delta - \alpha) + k \delta$ with $k > 0$. To this end, we use the equations
\begin{gather}
e_{\alpha + k \delta} = [2]_q^{-1} (e_{\alpha + (k - 1) \delta} \, e'_\delta - e'_\delta \, e_{\alpha + (k - 1) \delta}), \label{eapkd} \\
e_{(\delta - \alpha) + k \delta} = [2]_q^{-1} (e'_\delta \, e_{(\delta - \alpha) + (k - 1) \delta} - e_{(\delta - \alpha) + (k - 1) \delta} \, e'_\delta). \label{edmapkd}
\end{gather}
Finally, we construct root vectors corresponding to the roots $k \delta$ with $k > 1$ by the relations
\begin{equation}
e'_{k \delta} = e_{\alpha + (k - 1) \delta} \, e_{\delta - \alpha} - q^{-2} e_{\delta - \alpha} \, e_{\alpha + (k - 1) \delta}. \label{epkd}
\end{equation}
The redefinition of the vectors $e'_{k \delta}$ mentioned above is performed with the help of the relation
\begin{equation}
\kappa_q e_\delta(\zeta) = \log(1 + \kappa_q e'_\delta(\zeta)), \label{edepd}
\end{equation}
where
\begin{equation*}
e'_\delta(\zeta) = \sum_{k = 1}^\infty e'_{k \delta} \, \zeta^k, \qquad e_\delta(\zeta) = \sum_{k = 1}^\infty e_{k \delta} \, \zeta^k.
\end{equation*}
To construct root vectors corresponding to the negative roots we start with
\begin{equation*}
f_\alpha = f_1, \qquad f_{\delta - \alpha} = f_0.
\end{equation*}
Then, with the help of
\begin{equation}
f'_\delta = f_{\delta - \alpha} \, f_\alpha - q^2 f_\alpha \, f_{\delta - \alpha}, \label{f2}
\end{equation}
we define
\begin{gather}
f_{\alpha + k \delta} = [2]_q^{-1} (f'_\delta \, f_{\alpha + (k - 1) \delta} - f_{\alpha + (k - 1) \delta} \, f'_\delta), \label{fapkd} \\
f_{(\delta - \alpha) + k \delta} = [2]_q^{-1} (f_{(\delta - \alpha) + (k - 1) \delta} \, f'_\delta - f'_\delta \, f_{(\delta - \alpha) + (k - 1) \delta}, \label{fdmapkd}
\end{gather}
and 
\begin{equation}
f'_{k \delta} = f_{\delta - \alpha} \, f_{\alpha + (k - 1) \delta} - q^2 f_{\alpha + (k - 1) \delta} \, f_{\delta - \alpha}. \label{fpkd}
\end{equation}
The last step is to redefine the vectors $f_{k \delta}$ with the help of the relation 
\begin{equation}
{} - \kappa_q f_\delta(\zeta) = \log(1 - \kappa_q f'_\delta(\zeta)), \label{fdfpd}
\end{equation}
where
\begin{equation*}
f'_\delta(\zeta) = \sum_{k = 1}^\infty f'_{k \delta} \, \zeta^{- k}, \qquad f_\delta(\zeta) = \sum_{k = 1}^\infty f_{k \delta} \, \zeta^{- k}.
\end{equation*}

\subsubsection{Universal $R$-matrix} \label{sss:urm}

As any Hopf algebra the quantum group $\uqlslii$ has another comultiplication called the opposite comultiplication. It is defined by the equation
\begin{equation}
\Delta^\op = \Pi \circ \Delta, \label{deltaop}
\end{equation}
where
\begin{equation*}
\Pi(a \otimes b) = b \otimes a
\end{equation*}
for all $a, b \in \uqlslii$. When the quantum group $\uqlslii$ is defined as a $\bbC[[\hbar]]$-algebra it is a quasitriangular Hopf algebra. It means that there exists an element $\calR \in \uqlslii \otimes \uqlslii$, called the universal $R$-matrix, such that
\begin{equation}
\Delta^\op(a) = \calR \, \Delta(a) \, \calR^{-1} \label{urmd}
\end{equation}
for all $a \in \uqlslii$, and\footnote{For the explanation of the notation see, for example, the book \cite{ChaPre94} or the papers \cite{BooGoeKluNirRaz10, BooGoeKluNirRaz14a}.}
\begin{equation}
(\Delta \otimes \id) (\calR) = \calR^{13} \calR^{23}, \qquad (\id \otimes \Delta) (\calR) = \calR^{13} \calR^{12}. \label{urm}
\end{equation}
The most important property of the universal $R$-matrix is the equality
\begin{equation}
\calR^{12} \calR^{13} \calR^{23} = \calR^{23} \calR^{13} \calR^{12} \label{rrr}
\end{equation}
called the Yang--Baxter equation for the universal $R$-matrix.

The expression for the universal $R$-matrix of $\uqlslii$ considered as a $\bbC[[\hbar]]$-algebra can be constructed using the procedure proposed by Khoroshkin and Tolstoy \cite{TolKho92}. Note that here the universal $R$-matrix is an element of $\uqbp \otimes \uqbm$, where $\uqbp$ is the Borel subalgebra of $\uqlslii$ generated by $e_i$, $i = 0, 1$, and $q^x$, $x \in \widetilde{\gothh}$, and $\uqbm$ is the Borel subalgebra of $\uqlslii$ generated by $f_i$, $i = 0, 1$, and $q^x$, $x \in \widetilde{\gothh}$.

In fact, one can use the expression for the universal $R$-matrix from the paper \cite{TolKho92} also for the case of the quantum group $\uqlslii$ defined as a $\bbC$-algebra having in mind that in this case the quantum group is quasitriangular only in some restricted sense. Namely, all the relations involving the universal $R$-matrix should be considered as valid only for the weight representations of $\uqlslii$, see in this respect the paper \cite{Tan92}. This means that for any pair $\varphi$ and $\psi$ of weight representations of $\uqlslii$ on the vector spaces $V$ and $U$ one can define the element $\calR_{\varphi, \psi} \in \End(V) \otimes \End(U)$ satisfying the relations which allow to work with it as with an image of a real universal $R$-matrix. One can generalise the approach of the paper \cite{Tan92} to the case when $\varphi$ is an arbitrary homomorphism from $\uqlslii$ to an algebra $\calA$. In this case $\calR_{\varphi, \psi}$ is an element of $\calA \otimes \End(U)$ which is constructed in the following way.

Let $\psi$ be a weight representation of $\uqlslii$ on the vector space $U$. This means that
\begin{equation*}
U = \bigoplus_{\lambda \in \widetilde \gothh^*} U_\lambda,
\end{equation*}
where
\begin{equation*}
U_\lambda = \{u \in U \mid q^x u = q^{\lambda(x)} u \mbox{ for any } x \in \widetilde \gothh \}.
\end{equation*}
Taking into account relations (\ref{qh0h1}) and (\ref{c}), we conclude that $U_\lambda \ne \{0\}$ only if
\begin{equation}
\lambda(h_0 + h_1) = 0. \label{lambdah0h1h2}
\end{equation}
Let $\varphi$ be a homomorphism from $\uqlslii$ to some algebra $\calA$. Define the element $\calR_{\varphi, \psi} \in \calA \otimes \End(U)$ as
\begin{equation}
\calR_{\varphi, \psi} = (\varphi \otimes \psi)(\calR_{\prec \delta} \, \calR_{\sim \delta} \, \calR_{\succ \delta}) \, \calK_{\varphi, \psi}. \label{rpipi}
\end{equation}
The factor $\calR_{\prec \delta}$ is the product over $k \in \bbZ_+$ of the $q$-exponentials
\begin{equation*}
\calR_{\alpha + k \delta} = \exp_{q^{-2}} (\kappa_q \, e_{\alpha + k \delta} \otimes f_{\alpha + k \delta}).
\end{equation*}
Here and below we use the $q$-exponential defined as
\begin{equation*}
\exp_q(x) = \sum_{n = 0}^\infty q^{- n (n - 1) / 4} \frac{x^n}{[n]_q^{1/2}!} 
\end{equation*}
with
\begin{equation*}
[n]_q! = [n]_q [n - 1]_q \ldots [1]_q.
\end{equation*}
The order of the factors in $\calR_{\prec \delta}$ coincides with the chosen normal order of the roots $\alpha + k \delta$. For the factor $\calR_{\sim \delta}$ we have
\begin{equation}
\calR_{\sim \delta} = \exp \biggl(\kappa_q \sum_{k = 1}^\infty e_{k \delta} \otimes f_{k \delta} \biggr). \label{rsd}
\end{equation}
The last factor $\calR_{\succ \delta}$ is the product over $k \in \bbZ_+$ of the $q$-exponentials
\begin{equation*}
\calR_{(\delta - \alpha) + k \delta} = \exp_{q^{-2}} \bigl(\kappa_q \, e_{(\delta - \alpha) + k \delta} \otimes f_{(\delta - \alpha) + k \delta}\bigr).
\end{equation*}
The order of the factors in $\calR_{\succ \delta}$ coincides with the chosen normal order of the roots $(\delta - \alpha) + k \delta$.

To define the element $\calK_{\varphi, \psi}$  assume that $\{u_i\}$ is a basis of $U$ consisting of weight vectors and $\{E_{ij}\}$ is the corresponding basis of $\End(U)$. Now, if
\begin{equation*}
q^x u_i = q^{\lambda_i(x)} u_i
\end{equation*}
then
\begin{equation}
\calK_{\varphi, \psi} = \sum_i \varphi(q^{h_1 \lambda_i(h_1) / 2}) \otimes E_{ii}. \label{kpipii}
\end{equation}
It follows from (\ref{lambdah0h1h2}) that (\ref{kpipii}) can be written in a more symmetric form
\begin{equation}
\calK_{\varphi, \psi} = \sum_i \varphi(q^{(h_0 \lambda_i(h_0) + h_1 \lambda_i(h_1)) / 4}) \otimes E_{ii}. \label{kpipi}
\end{equation}
One can show that it is possible to work with $\calR_{\varphi, \psi}$ defined by (\ref{rpipi}) as if it is an image of some object which satisfies the same relations as the corresponding universal $R$-matrix should satisfy.

\subsection{Universal monodromy and universal transfer operators}

\subsubsection{General remarks} \label{ss:gr}

To construct integrability objects we have to endow $\uqlslii$ with a $\bbZ$-gradation, see, for example, \cite{BooGoeKluNirRaz14a, BooGoeKluNirRaz13}. The usual way to do it is as follows. Given $\zeta \in \bbC^\times$, we define an automorphism $\Gamma_\zeta$ of $\uqlslii$ by its action on the generators of $\uqlslii$ as
\begin{equation*}
\Gamma_\zeta(q^x) = q^x, \qquad \Gamma_\zeta(e_i) = \zeta^{s_i} e_i, \qquad \Gamma_\zeta(f_i) = \zeta^{-s_i} f_i,
\end{equation*}
where $s_i$ are arbitrary integers. The family of automorphisms $\Gamma_\zeta$, $\zeta \in \bbC^\times$, generates the $\bbZ$-gradation with the grading subspaces
\begin{equation*}
\uqlslii_m = \{ a \in \uqlslii \mid \Gamma_\zeta(a) = \zeta^m a \}.
\end{equation*}
Taking into account (\ref{cmul}), we see that
\begin{equation}
(\Gamma_\zeta \otimes \Gamma_\zeta) \circ \Delta = \Delta \circ \Gamma_\zeta. \label{ggd}
\end{equation}
It also follows from the explicit form of the universal $R$-matrix obtained with the help of the Tolstoy--Khoroshkin construction that for any $\zeta \in \bbC^\times$ we have
\begin{equation}
(\Gamma_\zeta \otimes \Gamma_\zeta)(\calR) = \calR.  \label{ggr}
\end{equation}
Following the physics tradition, we call $\zeta$ the spectral parameter.

Let $\varphi$ be a homomorphism of $\uqlslii$ to some unital associative algebra $\calA$. Given $\zeta \in \bbC^\times$, we denote by $\varphi_\zeta$ the homomorphism of $\uqlslii$ to $\calA$ given by the equation
\begin{equation}
\varphi_\zeta = \varphi \circ \Gamma_\zeta. \label{vpz}
\end{equation}
In particular, $\varphi$ can be a representation of $\uqlslii$ on a vector space $V$. In this case $\calA = \End(V)$.
If we consider $V$ as a $\uqlslii$-module corresponding to a representation $\varphi$, we denote by $V_\zeta$ the $\uqlslii$-module corresponding to the representation $\varphi_\zeta$. Certainly, as vector spaces $V$ and $V_\zeta$ coincide.

The universal monodromy operator $\calM_\varphi(\zeta)$ corresponding to the homomorphism $\varphi$ is defined by the relation
\begin{equation*}
\calM_\varphi(\zeta) = (\varphi_\zeta \otimes \id)(\calR).
\end{equation*}
It is clear that $\calM_\varphi(\zeta)$ is an element of $\calA \otimes \uqlslii$.

Universal monodromy operators are auxiliary objects needed for the construction of universal transfer operators. The universal transfer operator $\calT_\varphi(\zeta)$ corresponding to the universal monodromy operator $\calM_\varphi(\zeta)$ is defined as
\begin{equation*}
\calT_\varphi(\zeta) = (\tr_\calA \otimes \id)(\calM_\varphi(\zeta) (\varphi_\zeta(t) \otimes 1)) = ((\tr_\calA \circ \varphi_\zeta) \otimes \id)(\calR (t \otimes 1)),
\end{equation*}
where $t$ is a group-like element of $\uqlslii$ called a twist element, and $\tr_\calA$ is a trace on the algebra $\calA$ which means that $\tr_\calA$ is a linear mapping from $\calA$ to $\bbC$ satisfying the cyclicity condition
\begin{equation*}
\tr_\calA(a b) = \tr_\calA(b a)
\end{equation*}
for all $a, b \in \calA$. In particular, if $\pi$ is a representation of $\calA$ then $\tr \circ \pi$, where $\tr$ is a usual operator trace, is a trace on $\calA$. Note also that $\tr_\calA \circ \varphi$ is a trace on $\uqlslii$.

It is clear that $\calT_\varphi(\zeta)$ is an element of $\uqlslii$. An important property of the universal transfer operators $\calT_\varphi(\zeta)$ is that they commute for all homomorphisms $\varphi$ and all values of $\zeta$. They also commute with all generators $q^x$, $x \in \widetilde \gothh$, see, for example, our papers \cite{BooGoeKluNirRaz14a, BooGoeKluNirRaz13}. 

\subsubsection{Jimbo's homomorphism and universal monodromy operators}

As we noted above, for the case of the quantum group $\uqlslii$ the most important homomorphism used to construct integrability objects is the Jimbo's homomorphism. It is a homomorphism $\varphi: \uqlslii \to \uqglii$ defined by the equations\footnote{Recall that the Cartan generator $H$ of $\gothslii$ is related to the Cartan generators of $\gothglii$ by the relation $H = G_1 - G_2$.}
\begin{align}
& \varphi(q^{\nu h_0}) = q^{\nu(G_2 - G_1)}, && \varphi(q^{\nu h_1}) = q^{\nu (G_1 - G_2)}, \label{jha} \\*
& \varphi(e_0) = F \, q^{- G_1 - G_2}, && \varphi(e_1) = E, \\*
& \varphi(f_0) = E \, q^{G_1 + G_2} , && \varphi(f_1) = F, \label{jhc}
\end{align}
see the paper \cite{Jim86a}. Now, starting with the infinite dimensional representations $\widetilde \pi^\lambda$ of $\uqglii$, we define the infinite dimensional representations
\begin{equation}
\widetilde \varphi^\lambda = \widetilde \pi^\lambda \circ \varphi, \label{tfl}
\end{equation}
of $\uqlslii$, and denote the corresponding universal monodromy operators as $\widetilde \calM^\lambda(\zeta)$. Similarly, starting with the finite dimensional representations $\pi^\lambda$, we define the finite dimensional representations
\begin{equation*}
\varphi^\lambda = \pi^\lambda \circ \varphi,
\end{equation*}
and denote the corresponding universal monodromy operators as $\calM^\lambda(\zeta)$. Slightly abusing notation, we denote the $\uqlslii$-modules corresponding to the representations $\widetilde \varphi{}^\lambda$ and $\varphi^\lambda$ by $\widetilde V{}^\lambda$ and $V^\lambda$. For these modules we have
\begin{align}
& q^{\nu h_0} v_n = q^{\nu(- \mu + 2 n)} v_n, & &  q^{\nu h_1} v_n = q^{\nu(\mu - 2 n)} v_n, \label{vlh} \\*
& e_0 v_n = \zeta^{s_0} q^{- \lambda_1 - \lambda_2} v_{n + 1}, & & e_1 v_n = \zeta^{s_1} [\mu - n + 1]_q [n]_q v_{n - 1}, \label{vle} \\*
& f_0 v_n = \zeta^{-s_0} q^{\lambda_1 + \lambda_2} [\mu - n + 1]_q [n]_q v_{n - 1}, & & f_1 v_n = \zeta^{- s_1} v_{n + 1}. \label{vlf}
\end{align}

Additional universal monodromy operators can be defined with the help of the automorphism $\sigma$ of $\uqlslii$ defined by the relations
\begin{align}
& \sigma(q^{\nu h_0}) = q^{\nu h_1}, && \sigma(q^{\nu h_1}) = q^{\nu h_0}, \label{sigmah} \\*
& \sigma(e_0) = e_1, && \sigma(e_1) = e_0, \\*
& \sigma(f_0) = f_1, && \sigma(f_1) = f_0. \label{sigmaf}
\end{align}
Using the automorphism $\sigma$, we define a family of homomorphisms from $\uqlslii$ to $\uqglii$ generalizing the Jimbo's homomorphism as
\begin{equation*}
\varphi_i = \varphi \circ \sigma^{- i + 1},
\end{equation*}
and the corresponding universal monodromy operators as
\begin{equation*}
\calM_i(\zeta) = ((\varphi_i)_\zeta \otimes \id)(\calR).
\end{equation*}
Since $\sigma^2$ is the identity automorphism of $\uqlslii$, we have
\begin{equation*}
\calM_{i + 2}(\zeta) = \calM_i(\zeta).
\end{equation*}
Therefore, there are only two different universal monodromy operators of such kind.

It follows from (\ref{cmul}) that
\begin{equation*}
(\sigma \otimes \sigma) \circ \Delta = \Delta \circ \sigma.
\end{equation*}
Similarly, (\ref{deltaop}) and (\ref{cmul}) give
\begin{equation*}
(\sigma \otimes \sigma) \circ \Delta^{\mathrm{op}} = \Delta^{\mathrm{op}} \circ \sigma.
\end{equation*}
Using the definition of the universal $R$-matrix (\ref{urmd}), we obtain the equation
\begin{equation*}
((\sigma \otimes \sigma)(\calR)) \Delta(\sigma(a)) ((\sigma \otimes \sigma)(\calR))^{-1} = \Delta^{\mathrm{op}}(\sigma(a)).
\end{equation*}
Taking into account the uniqueness theorem for the universal $R$-matrix \cite{KhoTol92}, we conclude that
\begin{equation}
(\sigma \otimes \sigma)(\calR) = \calR. \label{ssr}
\end{equation}
Using this relation, it is not difficult to demonstrate that
\begin{equation}
\calM_{i + 1}(\zeta) = (\id \otimes \sigma)(\calM_i(\zeta))|_{s \to \sigma(s)}, \label{mtom}
\end{equation}
where $s \to \sigma(s)$ stands for
\begin{equation*}
s_0 \to s_1, \ s_1 \to s_0.
\end{equation*}

\subsubsection{Universal transfer operators} \label{sss:uto}

To proceed to universal transfer operators we need to define traces on $\uqglii$. The common way to do this is to use representations of $\uqglii$. Starting with the infinite dimensional representations $\widetilde \pi^{\lambda}$ of the quantum group $\uqglii$ described in section \ref{sss:hwm}, we define the traces
\begin{equation*}
\widetilde \tr^\lambda = \tr \circ \widetilde \pi^{\lambda}.
\end{equation*}
Similarly, when $\lambda_1 - \lambda_2$ is a non-negative integer, using the finite dimensional representations $\pi^\lambda$, we define the traces
\begin{equation*}
\tr^\lambda = \tr \circ \pi^\lambda.
\end{equation*}

In the case of an infinite dimensional representation there is a problem of convergence which can be solved with the help of a nontrivial twist element. We use a twist element of the form
\begin{equation}
t = q^{(\phi_0 h_0 + \phi_1 h_1) / 4}, \label{t}
\end{equation}
where $\phi_0$ and $\phi_1$ are complex numbers. Taking into account (\ref{qh0h1})and (\ref{c}), we assume that
\begin{equation*}
\phi_0 + \phi_1 = 0.
\end{equation*}

Now, we define a family of universal transfer operators associated with the infinite dimensional representations $\widetilde \pi^\lambda$ of $\uqglii$ as
\begin{equation*}
\widetilde \calT_i^\lambda(\zeta) = (\widetilde \tr{}^\lambda \otimes \id)(\calM_i(\zeta) ((\varphi_i)_\zeta(t) \otimes 1)) = ((\widetilde \tr{}^\lambda \circ (\varphi_i)_\zeta) \otimes \id)(\calR (t \otimes 1)),
\end{equation*}
and a family of universal transfer operators associated with the finite dimensional representations $\pi^\lambda$ of $\uqglii$ as
\begin{equation}
\calT_i^\lambda(\zeta) = (\tr^\lambda \otimes \id)(\calM_i(\zeta) ((\varphi_i)_\zeta(t) \otimes 1)) = ((\tr^\lambda \circ (\varphi_i)_\zeta) \otimes \id)(\calR (t \otimes 1)). \label{fduto}
\end{equation}

Let us discuss the dependence of the universal transfer operators on the spectral parameter $\zeta$. Consider, for example, the universal transfer operator $\widetilde \calT^\lambda(\zeta)$. From the structure of the universal $R$-matrix, it follows that the dependence on $\zeta$ is determined by the dependence on $\zeta$ of the elements of the form $\varphi_\zeta(a)$, where $a \in \uqbp$. Any such element is a linear combination of monomials each of which is a product of $E$, $F$ and $q^X$ for some $X \in \gothg$. Let $A$ be such a monomial. We have
\begin{equation*}
q^H A q^{- H} = q^{2 (n_1 - n_2)} A,
\end{equation*}
where $n_1$ and $n_2$ are the numbers of $E$ and $F$ in $A$. Hence $\widetilde \tr{}^\lambda(A)$ can be non-zero only if
\begin{equation*}
n_1 = n_2 = n.
\end{equation*}
Each $E$ enters $A$ with the factor $\zeta^{s_1}$ and each $F$ with the factor $\zeta^{s_0}$. Thus, for a monomial with non-zero trace we have a dependence on $\zeta$ of the form $\zeta^{n s}$. Therefore, the universal transfer operator $\widetilde \calT^\lambda(\zeta)$ depends on $\zeta$ only via $\zeta^s$, where $s$ is the sum of $s_0$ and $s_1$. The same is evidently true for all other universal transfer operators defined above. Using this fact, we obtain from (\ref{mtom}) the relation
\begin{equation}
\widetilde \calT{}^\lambda_{i + 1}(\zeta) = \sigma(\widetilde \calT^\lambda_i(\zeta))|_{\phi \to \sigma(\phi)}, \label{tst}
\end{equation}
where $\phi \to \sigma(\phi)$ stands for
\begin{equation*}
\phi_0 \to \phi_1, \qquad \phi_1 \to \phi_0.
\end{equation*}
Similarly, for the universal transfer operators corresponding to the finite dimensional representations $\pi^\lambda$ we have
\begin{equation}
\calT{}^\lambda_{i + 1}(\zeta) = \sigma(\calT^\lambda_i(\zeta))|_{\phi \to \sigma(\phi)}. \label{tsta}
\end{equation}

In the case when $\lambda_1 - \lambda_2$ is a non-negative integer, it follows from the exact sequence (\ref{fdm}) that 
\begin{equation*}
\widetilde \tr{}^{(\lambda_1, \lambda_2)} = \tr^{(\lambda_1, \lambda_2)} + \widetilde \tr{}^{(\lambda_2 - 1, \lambda_1 + 1)}.
\end{equation*}
Hence, in this case
\begin{equation*}
\tr^{(\lambda_1, \lambda_2)} = \widetilde \tr{}^{(\lambda_1, \lambda_2)} - \widetilde \tr{}^{(\lambda_2 - 1, \lambda_1 + 1)}.
\end{equation*}
We consider the above equation as the definition of the trace $\tr^\lambda$ and equation (\ref{fduto}) as the definition of the transfer operators $\calT_i^\lambda(\zeta)$ for an arbitrary $\lambda \in \gothg^*$. It is clear that we have
\begin{equation}
\calT^{(\lambda_1, \lambda_2)}_i(\zeta) = \widetilde \calT{}^{(\lambda_1, \lambda_2)}_i(\zeta) - \widetilde \calT{}^{(\lambda_2 - 1, \lambda_1 + 1)}_i(\zeta). \label{tetmt}
\end{equation}
It follows from this relation that
\begin{equation*}
\calT^{(\lambda_2 - 1, \, \lambda_1 + 1)}_i(\zeta) = - \calT^{(\lambda_1, \, \lambda_2)}_i(\zeta).
\end{equation*}
In particular, we have
\begin{equation*}
\calT^{(\nu, \, \nu + 1)}_i(\zeta) = 0
\end{equation*}
for any $\nu \in \bbC$.

\subsection{\texorpdfstring{Universal $L$-operators and universal $Q$-operators}{Universal L-operators and universal Q-operators}}

\subsubsection{General remarks}

It follows from (\ref{rpipi}) and (\ref{kpipi}) that to construct universal monodromy operators and transfer operators it suffices to have a representation of the Borel subalgebra $\uqbp$. This observation is used to construct universal $L$-operators and $Q$-operators. In distinction to the case of universal transfer operators, we use here representations of $\uqbp$ which cannot be obtained by restriction of representations of $\uqlslii$, or equivalently, representations of $\uqbp$ which cannot be extended to representations of $\uqlslii$. It is clear that to have interesting functional relations one should use for the construction of universal $L$-operators and $Q$-operators representations which are connected in some way to the representations used for the construction of universal monodromy operators and transfer operators.

In general, a universal $L$-operator associated with a homomorphism $\rho$ from $\uqbp$ to some associative algebra $\calB$ is defined as
\begin{equation*}
\calL_\rho(\zeta) = (\rho_\zeta \otimes \id)(\calR).
\end{equation*}
As the universal monodromy operators are auxiliary objects needed for the construction of the universal transfer operators, the universal $L$-operators are needed for the construction of the universal $Q$-operators. The universal $Q$-operator $\calQ_\rho(\zeta)$ corresponding to the universal $L$-operator $\calL_\rho(\zeta)$ is defined as
\begin{equation*}
\calQ_\rho(\zeta) = (\tr_\calB \otimes \id)(\calL_\rho(\zeta) (\rho_\zeta(t) \otimes 1)) = ((\tr_\calB \circ \rho_\zeta) \otimes \id)(\calR (t \otimes 1)),
\end{equation*}
where $\tr_\calB$ is a trace on $\calB$, and $t$ is a twist element.

\subsubsection{Basic representation} \label{sss:br}

We start the construction of universal $L$-operators and $Q$-ope\-ra\-tors with the construction of the basic homomorphism from $\uqbp$. The initial point is the infinite dimensional representations $\widetilde \varphi^\lambda$ of $\uqlslii$ defined by equation (\ref{tfl}).

First define the notion of a shifted representation. Let $\xi$ be an element of $\widetilde \gothh{}^*$ satisfying the equation
\begin{equation*}
\xi(h_0 + h_1) = 0.
\end{equation*}
If $\varphi$ is a representation of $\uqbp$, then the representation $\varphi[\xi]$ defined by the relations
\begin{equation*}
\varphi[\xi](e_i) = \varphi(e_i), \qquad \varphi[\xi](q^x) = q^{\xi(x)} \varphi (q^x)
\end{equation*}
is a representation of $\uqbp$ called a shifted representation. If $V$ is a $\uqbp$-module corresponding to the representation $\varphi$, then $V[\xi]$ denotes the $\uqbp$-module corresponding to the representation $\varphi[\xi]$.

Consider the restriction of the representation $\widetilde \varphi^\lambda$ to $\uqbp$ and a shifted representation $\widetilde \varphi^\lambda[\xi]$ of $\uqbp$. One can show that for a non-zero $\xi$ this representation cannot be extended to a representation of $\uqlslii$ and we can use it to construct a universal $Q$-operator. However, it follows from (\ref{rpipi}) and (\ref{kpipi}) that the universal $Q$-operator defined with the help of the representation $\widetilde \varphi^\lambda[\xi]$ is connected with the universal transfer operator $\widetilde \calT{}^\lambda(\zeta)$ defined with the help of the representation $\widetilde \varphi^\lambda$ by the relation
\begin{equation}
\calQ_{\widetilde \varphi^\lambda[\xi]}(\zeta) = \widetilde \calT{}^\lambda(\zeta) \, q^{(\xi(h^{\mathstrut}_0) h'_0 + \xi(h^{\mathstrut}_1) h'_1)/ 4}, \label{shrtr}
\end{equation}
where
\begin{equation*}
h'_i = h_i + \phi_i.
\end{equation*}
Here we assume that the twist element is of the form (\ref{t}). We see that the use of shifted representations does not give anything really new.

Consider another method to obtain a new representation of $\uqbp$. The restriction of the representation $(\widetilde \varphi^\lambda)_\zeta$ to $\uqbp$ is described by equations (\ref{vlh}) and (\ref{vle}). Let us try to go to the limit $\mu \to \infty$. Looking at relations (\ref{vlh}) and (\ref{vle}), we see that we cannot perform this limit directly. Therefore, we consider first a shifted representation $(\widetilde \varphi^\lambda)_\zeta[\xi]$ with $\xi$ defined by the relations
\begin{equation*}
\xi(h_0) = \mu, \qquad \xi(h_1) = - \mu.
\end{equation*}
Then we introduce a new basis
\begin{equation*}
w_n = c^n v_n,
\end{equation*}
where
\begin{equation*}
c = q^{- \mu - 1 - (2 \lambda_2 - 1) s_1 / s}.
\end{equation*}
Now relations (\ref{vlh}) take the form
\begin{equation}
q^{\nu h_0} w_n = q^{2 \nu n} w_n, \qquad q^{\nu h_1} w_n = q^{- 2 \nu n} w_n, \label{bhwn}
\end{equation}
and instead of relations (\ref{vle}) we have
\begin{equation}
e_0 w_n = \widetilde \zeta^{s_0} w_{n + 1}, \qquad e_1 w_n = \widetilde \zeta^{s_1} \kappa_q^{-1} (q^{- n} - q^{- 2 \mu + n - 2}) [n]_q w_{n - 1}, \label{bewn}
\end{equation}
where
\begin{equation*}
\widetilde \zeta = q^{- (2 \lambda_2 - 1) / s} \zeta.
\end{equation*}
It is possible now to consider the limit $\mu \to \infty$. Relations (\ref{bhwn}) retain their form, while (\ref{bewn}) go to
\begin{equation*}
e_0 w_n = \widetilde \zeta^{s_0} w_{n + 1}, \qquad e_1 w_n = \widetilde \zeta^{s_1} \kappa_q^{-1} q^{- n} [n]_q w_{n - 1}.
\end{equation*}

Denote by $\theta$ the representation of $\uqbp$ defined by the relations
\begin{align}
& q^{\nu h_0} v_n = q^{2 \nu n} v_n, & & q^{\nu h_1} v_n = q^{- 2 \nu n} v_n, \label{qhv} \\
& e_0 v_n = v_{n + 1}, & & e_1 v_n = \kappa_q^{-1} q^{- n} [n]_q v_{n - 1}, \label{evb}
\end{align}
and by $W$ the corresponding $\uqbp$-module. It is clear that if we define the universal $Q$-operator $\calQ'(\zeta)$ by the relation
\begin{equation*}
\calQ'(\zeta) = ((\tr \circ \theta_\zeta) \otimes \id)(\calR (t \otimes 1)),
\end{equation*}
then, having in mind (\ref{shrtr}), we obtain
\begin{equation*}
\calQ'(\zeta) = \lim_{\mu \to \infty} \left( \widetilde \calT^{(\mu, 0)}(q^{- 1 / s} \zeta) \, q^{\mu (h'_0 - h'_1) / 4} \right).
\end{equation*}
Here the prime means that we will redefine $Q$-operators. It follows from the above relation that the universal $Q$-operators $\calQ'(\zeta)$ for all values of $\zeta$ commute. In addition, they commute with all universal transfer operators $\widetilde \calT{}_i^\lambda(\zeta)$, $\calT_i^\lambda(\zeta)$ and with all generators $q^x$, $x \in \widetilde \gothh$,  see, for example, our papers \cite{BooGoeKluNirRaz14a, BooGoeKluNirRaz13}.

We use the representation $\theta$ as the basic representation for the construction of all necessary universal $L$-operators and $Q$-operators. In fact, it is an asymptotic, or prefundamental, representation of $\uqbp$, see the papers  \cite{HerJim12, FreHer13}.

\subsubsection{Interpretation in terms of $q$-oscillators} \label{sss:ito}

It is useful to give an interpretation of relations (\ref{qhv})--(\ref{evb}) in terms of $q$-oscillators. Let us remind the necessary definitions, see, for example, the book \cite{KliSch97}.

Let $\hbar$ be a complex number and $q = \exp \hbar$.\footnote{We again assume that $q$ is not a root of unity.} The $q$-oscillator algebra $\Osc_q$ is a unital associative $\bbC$-algebra with generators $b^\dagger$, $b$, $q^{\nu N}$, $\nu \in \bbC$, and relations
\begin{gather*}
q^0 = 1, \qquad q^{\nu_1 N} q^{\nu_2 N} = q^{(\nu_1 + \nu_2)N}, \\
q^{\nu N} b^\dagger q^{-\nu N} = q^\nu b^\dagger, \qquad q^{\nu N} b q^{-\nu N} = q^{-\nu} b, \\
b^\dagger b = [N]_q, \qquad b b^\dagger = [N + 1]_q,
\end{gather*}
where we use the notation similar to (\ref{xnq}). It is easy to understand that the monomials $(b^\dagger)^{k + 1} q^{\nu N}$, $b^{k + 1} q^{\nu N}$
and $q^{\nu N}$ for $k \in \bbZ_+$ and $\nu \in \bbC$ form a basis of $\Osc_q$.

Two representations of $\Osc_q$ are interesting for us. First, let $W^{\scriptscriptstyle +}$ be a free vector space generated by the set $\{ v_0, v_1, \ldots \}$. One can show that the relations
\begin{gather*}
q^{\nu N} v_n = q^{\nu n} v_n, \\*
b^\dagger v_n = v_{n + 1}, \qquad b \, v_n = [n]_q v_{n - 1},
\end{gather*}
where we assume that $v_{-1} = 0$, endow $W^{\scriptscriptstyle +}$ with the structure of an $\Osc_q$-module. We denote the corresponding representation of the algebra $\Osc_q$ by $\chi^{\scriptscriptstyle +}$. Further, let $W^{\scriptscriptstyle -}$ be a free vector space generated again by the set $\{ v_0, v_1, \ldots \}$. The relations
\begin{gather*}
q^{\nu N} v_n = q^{- \nu (n + 1)} v_n, \\
b \, v_n = v_{n + 1}, \qquad b^\dagger v_n = - [n]_q v_{n - 1},
\end{gather*}
where we again assume that $v_{-1} = 0$, endow the vector space $W^{\scriptscriptstyle -}$ with the structure of an $\Osc_q$-module. We denote the corresponding representation of $\Osc_q$ by $\chi^{\scriptscriptstyle -}$.

Assume that the generators of $\Osc_q$ act on the module $W$ defined by equations (\ref{qhv})--(\ref{evb}) as on the module $W^{\scriptscriptstyle +}$. This allows us to write (\ref{qhv})--(\ref{evb})  as
\begin{align*}
& q^{\nu h_0} v_n = q^{2 \nu N} v_n, && q^{\nu h_1} v_n = q^{- 2\nu N} v_n, \\*
& e_0 v_n = b^\dagger v_n, && e_1 v_n = \kappa_q^{-1} b \, q^{- N} v_n.
\end{align*}
These equations suggest defining a homomorphism $\rho: \uqbp \to \Osc_q$ by
\begin{align}
& \rho(q^{\nu h_0}) = q^{2 \nu N}, && \rho(q^{\nu h_1}) = q^{- 2\nu N}, \label{rhoh} \\*
& \rho(e_0) = b^\dagger, && \rho(e_1) = \kappa_q^{-1} b \, q^{- N}. \label{rhoe}
\end{align}
Using this homomorphism, we can write for the representation $\theta$ the equation
\begin{equation*}
\theta = \chi^{\scriptscriptstyle +} \circ \rho.
\end{equation*}

\subsubsection{Universal $L$-operators and universal $Q$-operators}

Define the universal $L$-operator\footnote{The prime here and below means that the corresponding universal $L$-operators will be used to define primed universal $Q$-operators.}
\begin{equation*}
\calL'_\rho(\zeta) = (\rho_\zeta \otimes \id)(\calR)
\end{equation*}
being an element of $\Osc_q \otimes \uqlslii$. Using the homomorphism $\rho$ defined by (\ref{rhoh})--(\ref{rhoe}) and the automorphism $\sigma$ defined by (\ref{sigmah})--(\ref{sigmaf}), we define the homomorphisms
\begin{equation*}
\rho_i = \rho \circ \sigma^{- i},
\end{equation*}
where $i = 1, 2$, and the universal $L$-operators
\begin{equation*}
\calL'_i(\zeta) = ((\rho_i)_\zeta \otimes \id)(\calR).
\end{equation*}
These universal $L$-operators are again elements of $\Osc_q \otimes \uqlslii$. In the same way as for the universal monodromy operators, using relation (\ref{ssr}), we obtain the equation
\begin{equation}
\calL'_{i + 1}(\zeta) = (\id \otimes \sigma)(\calL'_i(\zeta))|_{s \to \sigma(s)}. \label{lipo}
\end{equation}

To define universal $Q$-operators one should define traces on the algebra $\Osc_q$. Two standard representations $\chi^{\scriptscriptstyle +}$ and $\chi^{\scriptscriptstyle -}$ of $\Osc_q$ generate two traces. We denote
\begin{equation*}
\tr^{\scriptscriptstyle +} = \tr \circ \chi^{\scriptscriptstyle +}, \qquad \tr^{\scriptscriptstyle -} = \tr \circ \chi^{\scriptscriptstyle -}.
\end{equation*}
We see that
\begin{equation*}
\tr^{\scriptscriptstyle +} ((b^\dagger)^{k + 1} q^{\nu N}) = 0, \qquad \tr^{\scriptscriptstyle +} (b^{k + 1} q^{\nu N}) = 0,
\end{equation*}
and that
\begin{equation*}
\tr^{\scriptscriptstyle +}(q^{\nu N}) = (1 - q^\nu)^{-1}
\end{equation*}
for $|q| < 1$. For $|q| > 1$ we define the trace $\tr^{\scriptscriptstyle +}$ by analytic continuation. Since the monomials $(b^\dagger)^{k + 1} q^{\nu N}$, $b^{k + 1} q^{\nu N}$ and $q^{\nu N}$ for $k \in \bbZ_+$ and $\nu \in \bbC$ form a basis of $\Osc_q$, the above relations are enough to determine the trace of any element of $\Osc_q$. It appears that
\begin{equation}
\tr^{\scriptscriptstyle -} = - \tr^{\scriptscriptstyle +}. \label{trtr}
\end{equation}
Therefore, we will use only the trace $\tr^{\scriptscriptstyle +}$, and define the universal $Q$-operators corresponding to the universal $L$-operators $\calL'_i(\zeta)$ as
\begin{equation*}
\calQ'_i(\zeta) = (\tr^{\scriptscriptstyle +} \otimes \id) (\calL'_i(\zeta) ((\rho_i)_\zeta(t) \otimes 1)) = ((\tr^{\scriptscriptstyle +} \circ (\rho_i)_\zeta) \otimes \id) (\calR (t \otimes 1)).
\end{equation*}

Similarly as for the case of the universal transfer operators, one can demonstrate that the universal $Q$-operators $\calQ'_i(\zeta)$ depend on $\zeta$ via $\zeta^s$. Therefore, equation (\ref{lipo}) leads to the relation
\begin{equation}
\calQ'_{i + 1}(\zeta) = \sigma(\calQ'_i(\zeta))|_{\phi \to \sigma(\phi)}. \label{qsq}
\end{equation}

Based on section \ref{sss:br}, all universal $Q$-operators $\calQ'_i(\zeta)$ can be considered as limits of the corresponding universal transfer operators. Therefore, they commute for all values of $i$ and $\zeta$. They commute also with all universal transfer operators $\widetilde \calT{}_i^\lambda(\zeta)$, $\calT_i^\lambda(\zeta)$ and with all generators $q^x$, $x \in \widetilde \gothh$.

\section{Universal functional relations}

\subsection{Factorized and determinant representations of transfer operators}

\subsubsection{General remarks}

Roughly speaking, functional relations for quantum integrable systems are some relations connecting products of integrability objects, such as universal transfer operators and universal $Q$-operators. Here the most important are relations representing universal transfer operators via products of universal $Q$-operators. To analyse products of universal $Q$-operators one uses the following fact. Let $\theta_1$ and $\theta_2$ be two representations of the Borel subalgebra $\uqbp$. Define the corresponding universal $Q$-operators,
\begin{equation*}
\calQ_{\theta_1}(\zeta_1) = ((\tr \circ (\theta_1)_{\zeta_1}) \otimes \id)(\calR (t \otimes 1)), \qquad \calQ_{\theta_2}(\zeta_2) = ((\tr \circ (\theta_2)_{\zeta_2}) \otimes \id)(\calR (t \otimes 1)).
\end{equation*}
Denote by $\theta_{\zeta_1, \, \zeta_2}$ the tensor product of the representations $(\theta_1)_{\zeta_1}$ and $(\theta_2)_{\zeta_2})$,
\begin{equation*}
\theta_{\zeta_1, \, \zeta_2} = (\theta_1)_{\zeta_1} \otimes_\Delta (\theta_2)_{\zeta_2} = ((\theta_1)_{\zeta_1} \otimes (\theta_2)_{\zeta_2}) \circ \Delta.
\end{equation*}
One can show that
\begin{equation*}
\calQ_{\theta_1}(\zeta_1) \calQ_{\theta_2}(\zeta_2) = ((\tr \circ \theta_{\zeta_1, \, \zeta_2}) \otimes \id)(\calR (t \otimes 1)),
\end{equation*}
see, for example, the paper \cite{BooGoeKluNirRaz14a}. Hence, to analyse the product of universal $Q$-operators, one should analyse the tensor product of the corresponding representations.

\subsubsection{Tensor product of oscillator representations}

Let us consider the product of the universal $Q$-operators $\calQ'_1(\zeta_1)$ and $\calQ'_2(\zeta_2)$ defined in section \ref{sss:ito}. It is easy to see that the universal $Q$-operators can be represented as
\begin{equation*}
\calQ'_i(\zeta) = ((\tr \circ (\chi^\ssplus_i)_\zeta) \otimes \id)(\calR (t \otimes 1)) = - ((\tr \circ (\chi^\ssminus_i)_\zeta) \otimes \id)(\calR (t \otimes 1)),
\end{equation*}
where
\begin{equation*}
\chi^\ssplus_i = \chi^\ssplus \circ \rho_i, \qquad \chi^\ssminus_i = \chi^\ssminus \circ \rho_i.
\end{equation*}
Here $\chi^\ssplus$ and $\chi^\ssminus$ are representations of $\Osc_q$ described in section \ref{sss:ito}. The mappings $\chi^\ssplus_i$ and $\chi^\ssminus_i$ are representations of $\uqbp$. Denote the corresponding $\uqbp$-modules by $W^\ssplus_i$ and $W^\ssminus_i$.

It is convenient to use for $\calQ_1(\zeta_1)$ and $\calQ_2(\zeta_2)$ the representations
\begin{equation*}
\calQ'_1(\zeta_1) = - ((\tr \circ (\chi^\ssminus_1)_{\zeta_1}) \otimes \id)(\calR (t \otimes 1)), \qquad \calQ'_2(\zeta_2) = ((\tr \circ (\chi^\ssplus_2)_{\zeta_2}) \otimes \id)(\calR (t \otimes 1)),
\end{equation*}
and write
\begin{equation*}
\calQ'_1(\zeta_1) \calQ'_2(\zeta_2) = \calQ'_2(\zeta_2) \calQ'_1(\zeta_1) = - ((\tr \circ ((\chi^\ssplus_2)_{\zeta_2} \otimes_\Delta (\chi^\ssminus_1)_{\zeta_1})) \otimes \id)(\calR (t \otimes 1)).
\end{equation*}

Consider now the representation $(\chi^\ssplus_2)_{\zeta_2} \otimes_\Delta (\chi^\ssminus_1)_{\zeta_1})$. The corresponding $\uqbp$-mo\-du\-le $(W^\ssplus_2)_{\zeta_2} \otimes_\Delta (W^\ssminus_1)_{\zeta_1}$ is also an $(\Osc_q \otimes \Osc_q)$-module. We use for the generators of the algebra $\Osc_q \otimes \Osc_q$ the following notation
\begin{align*}
& b^\dagger_A = b^\dagger \otimes 1, && b_A = b \otimes 1, && q^{\nu N_A} = q^{\nu N} \otimes 1, \\
& b^\dagger_B = 1 \otimes b^\dagger, && b_B = 1 \otimes b, && q^{\nu N_B} = 1 \otimes q^{\nu N}.
\end{align*}
Using the explicit form of the comultiplication in $\uqbp$ determined by equations (\ref{cmul}), we obtain for the action of the generators of $\uqbp$ on $(W^\ssplus_2)_{\zeta_2} \otimes_\Delta (W^\ssminus_1)_{\zeta_1}$ the following expressions
\begin{equation*}
q^{\nu h_0} v = q^{2 \nu (N_A - N_B)} v, \qquad q^{\nu h_1} v = q^{- 2 \nu (N_A - N_B)} v,
\end{equation*}
and
\begin{align*}
& e_0 v = (\zeta_2^{s_0} b^\dagger_A + \zeta_1^{s_0} \kappa_q^{-1} b^{\mathstrut}_B q^{- 2 N_A - N_B}) v, \\
& e_1 v = (\zeta_2^{s_1} \kappa_q^{-1} b^{\mathstrut}_A q^{- N_A} + \zeta_1^{s_1} b^\dagger_B q^{2 N_A}) v.
\end{align*}
A convenient for our purposes basis of $(W^\ssplus_2)_{\zeta_2} \otimes_\Delta (W^\ssminus_1)_{\zeta_1}$ is formed by the vectors
\begin{equation*}
w_{n, k} = (\zeta^{- s_0} q^{\lambda_1 + \lambda_2} e_0)^n (\zeta_1^{- s_1} b_B)^k w_0, \qquad n, k \in \bbZ_+,
\end{equation*}
where $w_0$ is a unique vector of $(W^\ssplus_2)_{\zeta_2} \otimes_\Delta (W^\ssminus_1)_{\zeta_1}$ satisfying the relations
\begin{equation*}
b^{\mathstrut}_A w_0 = 0, \qquad b^\dagger_B w_0 = 0,
\end{equation*}
and $\lambda_1$, $\lambda_2$ and $\zeta$ are additional parameters. Using formulas of the paper \cite{BooGoeKluNirRaz14a}, one can easily show that
\begin{align}
& q^{\nu h_0} w_{n, k} = q^{2 \nu (n + k + 1)} w_{n, k}, \label{qh0w} \\
& q^{\nu h_1} w_{n, k} = q^{- 2 \nu (n + k + 1)} w_{n, k}, \\
& e_0 w_{n, k} = \zeta^{s_0} q^{-\lambda_1 - \lambda_2} w_{n + 1, k}, \\
& e_1 w_{n, k} = \zeta^{s_1} \zeta^{-s} q^{\lambda_1 + \lambda_2} (q^{-n} \zeta_2^s - q^n \zeta_1^s) [n]_q w_{n - 1, k} - q^{2n} [k]_q w_{n, k - 1}. \label{e1w}
\end{align}
Assume now that
\begin{equation*}
\zeta_1 = q^{-2 (\lambda_1 + 1/2)/s} \zeta, \qquad \zeta_2 = q^{-2(\lambda_2 - 1/2)/s} \zeta,
\end{equation*}
and compare in this case (\ref{qh0w})--(\ref{e1w}) with (\ref{vlh}) and (\ref{vle}). We see that there is an increasing filtration
\begin{equation*}
\{0\} = ((W^\ssplus_2)_{\zeta_2} \otimes_\Delta (W^\ssminus_1)_{\zeta_1})_{-1} \subset ((W^\ssplus_2)_{\zeta_2} \otimes_\Delta (W^\ssminus_1)_{\zeta_1})_0 \subset ((W^\ssplus_2)_{\zeta_2} \otimes_\Delta (W^\ssminus_1)_{\zeta_1})_1 \subset \ldots
\end{equation*}
formed by the submodules
\begin{equation*}
((W^\ssplus_2)_{\zeta_2} \otimes_\Delta (W^\ssminus_1)_{\zeta_1})_m = \bigoplus_{k = 0}^m \bigoplus_{n = 0}^\infty \bbC \, w_{n, k} 
\end{equation*}
with the quotient modules
\begin{equation*}
((W^\ssplus_2)_{\zeta_2} \otimes_\Delta (W^\ssminus_1)_{\zeta_1})_m / ((W^\ssplus_2)_{\zeta_2} \otimes_\Delta (W^\ssminus_1)_{\zeta_1})_{m - 1} \simeq (\widetilde V^\lambda)[\xi_m]_\zeta.
\end{equation*}
Here the elements $\xi_m \in \tgothh^*$ are given by the equations
\begin{equation*}
\xi_m(h_0) = \lambda_1 - \lambda_2 + 2 m + 2, \qquad \xi_m(h_1) = - \lambda_1 + \lambda_2 - 2 m - 2.
\end{equation*}

\subsubsection{Factorized and determinant representations}

Using relation (\ref{shrtr}) and the results of the above section we can write
\begin{equation*}
\calQ'_1(q^{- 2(\lambda_1 + 1/2)/s} \zeta) \calQ'_2(q^{- 2(\lambda_2 - 1/2)/s} \zeta) = - \widetilde \calT^\lambda(\zeta) \sum_{m = 0}^\infty q^{(\xi_m(h_0) h'_0 + \xi_m(h_1) h'_1)/4}.
\end{equation*}
The summation gives
\begin{equation}
\sum_{m = 0}^\infty q^{(\xi_m(h_0) h'_0 + \xi_m(h_1) h'_1)/4} = q^{(\lambda_1 - \lambda_2 + 2)(h'_0 - h'_1)/4} (1 - q^{(h'_0 - h'_1)/2})^{-1}. \label{sshrp}
\end{equation}
Introduce the notation
\begin{equation*}
\calD_1 = (h'_0 - h'_1)s/8, \qquad \calD_2 = (h'_1 - h'_0)s/8,
\end{equation*}
and note that
\begin{equation*}
\calD_1 + \calD_2 = 0.
\end{equation*}
Now equation (\ref{sshrp}) can be written as
\begin{equation*}
\sum_{m = 0}^\infty q^{(\xi_m(h_0) h'_0 + \xi_m(h_1) h'_1)/4} = - q^{2 (\lambda_1 + 1/2) \calD_1/s + 2 (\lambda_2 - 1/2) \calD_2/s} (q^{2\calD_1/s} - q^{2\calD_2/s})^{-1}.
\end{equation*}
Redefining $Q$ operators $\calQ'_1(\zeta)$ and $\calQ'_2(\zeta)$ in the following way,
\begin{equation*}
\calQ_1(\zeta) = \zeta^{\calD_1} \calQ'_1(\zeta), \qquad \calQ_2(\zeta) = \zeta^{\calD_2} \calQ'_2(\zeta),
\end{equation*}
we come to the equation
\begin{equation*}
\calC \, \widetilde \calT^\lambda(\zeta) = \calQ_1(q^{- 2(\lambda_1 + 1/2)/s} \zeta) \calQ_2(q^{- 2(\lambda_2 - 1/2)/s} \zeta),
\end{equation*}
where
\begin{equation*}
\calC = (q^{2\calD_1/s} - q^{2\calD_2/s})^{-1}.
\end{equation*}
Now using equation (\ref{tetmt}) we obtain the determinant representation
\begin{equation}
\calC \, \calT^{(\lambda_1 - 1/2, \, \lambda_2 + 1/2)}(\zeta) = \det \big(\calQ_i(q^{-2\lambda_j/s} \zeta)\big)_{i, j = 1, 2}. \label{tqq}
\end{equation}
This representation allows us to obtain all functional relations and discover some interesting properties of the transfer operators.

First of all we show that the transfer operators $\calT^\lambda_1(\zeta)$ and $\calT^\lambda_2(\zeta)$ coincide. To the end we observe that
\begin{equation}
\sigma(\calD_i)|_{\phi \to \sigma(\phi)} = \calD_{i + 1}, \label{sded}
\end{equation}
and, therefore,
\begin{equation*}
\sigma(\calC)|_{\phi \to \sigma(\phi)} = - \calC.
\end{equation*}
Furthermore, equation (\ref{sded}) gives
\begin{equation*}
\calQ'_{i + 1}(\zeta) = \sigma(\calQ'_i(\zeta))|_{\phi \to \sigma(\phi)}.
\end{equation*}
Applying the automorphism $\sigma$ to the determinant representation (\ref{tqq}), we see that
\begin{equation*}
\calT^\lambda_1(\zeta) = \calT^\lambda_2(\zeta) = \calT^\lambda(\zeta).
\end{equation*}
Another interesting consequence of (\ref{tqq}) is the equation 
\begin{equation}
\calT^{(\lambda_1 + \nu, \, \lambda_2 + \nu)}(q^{2 \nu / s} \zeta) = \calT^{(\lambda_1, \, \lambda_2)}(\zeta) \label{tls}
\end{equation}
valid for any $\nu \in \bbC$. This equation means that we can use only the transfer operators of the form $\calT^{(\nu, \, 0)}(\zeta)$. However, it is not always convenient.

\subsection{Functional relations in terms of universal integrability objects}

\subsubsection{\texorpdfstring{$TQ$-relations}{TQ-relations}}

Let $\lambda_j$, $j = 1, \ldots, 3$, be arbitrary complex numbers. It is evident that there is a trivial identity
\begin{equation*}
\left| \begin{array}{ccc}
\calQ_1(q^{- 2 \lambda_1/s} \zeta) & \calQ_1(q^{- 2 \lambda_2/s} \zeta) & \calQ_1(q^{- 2 \lambda_3/s} \zeta) \\
\calQ_2(q^{- 2 \lambda_1/s} \zeta) & \calQ_2(q^{- 2 \lambda_2/s} \zeta) & \calQ_2(q^{- 2 \lambda_3/s} \zeta) \\
\calQ_k(q^{- 2 \lambda_1/s} \zeta) & \calQ_k(q^{- 2 \lambda_2/s} \zeta) & \calQ_k(q^{- 2 \lambda_3/s} \zeta)
\end{array}
\right| = 0,
\end{equation*}
for any $k = 1, 2$. Expanding the determinant over the third row, we obtain the relation
\begin{multline*}
\calT^{(\lambda_1 - 1/2, \, \lambda_2 + 1/2)}(\zeta) \calQ_k(q^{- 2 \lambda_3 / s} \zeta) \\- \calT^{(\lambda_1 - 1/2, \, \lambda_3 + 1/2)}(\zeta) \calQ_k(q^{- 2 \lambda_2 / s} \zeta) + \calT^{(\lambda_2 - 1/2, \, \lambda_3 + 1/2)}(\zeta) \calQ_k(q^{- 2 \lambda_1 / s} \zeta) = 0.
\end{multline*}
We call this equation the universal $TQ$-relation. Assuming that
\begin{equation*}
\lambda_1 = 1, \qquad \lambda_2 = 0, \qquad \lambda_3 = -1,
\end{equation*}
we obtain the relation
\begin{equation}
\calT^{(1/2, \, 1/2)}(\zeta) \calQ_k(q^{2 / s} \zeta) - \calT^{(1/2, \, -1/2)}(\zeta) \calQ_k(\zeta) + \calT^{(-1/2, \, -1/2)}(\zeta) \calQ_k(q^{- 2 / s} \zeta) = 0. \label{tqi}
\end{equation}
It follows from the structure of the universal $R$-matrix that $\calT^{(0, \, 0)}(\zeta) = 1$. Therefore, as follows from (\ref{tls}), we have
\begin{equation*}
\calT^{(\nu, \, \nu)}(\zeta) = 1
\end{equation*}
for any $\nu \in \bbC$. This property leads to a simpler form of (\ref{tqi}):
\begin{equation*}
\calT^{(1, \, 0)}(q^{1/s}\zeta) \calQ_k(\zeta) = \calQ_k(q^{2 / s} \zeta) + \calQ_k(q^{- 2 / s} \zeta) = 0. \label{tq}
\end{equation*}
This equation is an analogue of the famous Baxter's $TQ$ -relations in the form independent of the representation of the quantum group in the quantum space.

\subsubsection{\texorpdfstring{$TT$-relations}{TT-relations}}

Now consider another identity
\begin{equation}
\left| \begin{array}{ccc}
\calQ_1(q^{- 2 \lambda_1/s} \zeta) & \calQ_1(q^{- 2 \lambda_2/s} \zeta) & \calQ_1(q^{- 2 \lambda_3/s} \zeta) \\
\calQ_2(q^{- 2 \lambda_1/s} \zeta) & \calQ_2(q^{- 2 \lambda_2/s} \zeta) & \calQ_2(q^{- 2 \lambda_3/s} \zeta) \\
\calT^{(\lambda_1 - 1/2, \, \lambda_4 + 1/2)}(\zeta) & \calT^{(\lambda_2 - 1/2, \, \lambda_4 + 1/2)}(\zeta) & \calT^{(\lambda_3 - 1/2, \, \lambda_4 + 1/2)}(\zeta)
\end{array} \label{ttdet}
\right| = 0,
\end{equation}
where $\lambda_j$, $j = 1, \ldots, 4$, are again arbitrary complex numbers. To prove this identity we just observe that
\begin{equation*}
\calC \, \calT^{(\lambda_j - 1/2, \, \lambda_4 + 1/2)}(\zeta) = \calQ_1(q^{- 2 \lambda_j/s} \zeta) \calQ_2(q^{- 2 \lambda_4/s} \zeta) - \calQ_2(q^{- 2 \lambda_j/s} \zeta) \calQ_1(q^{- 2 \lambda_4/s} \zeta)
\end{equation*}
for any $j = 1, 2, 3$. Hence, the last row of the matrix in (\ref{ttdet}) is a linear combination of the first two rows, and the identity is true. Expanding the determinant in (\ref{ttdet}) over the last row, we come to the equation
\begin{multline}
\calT^{(\lambda_1 - 1/2, \, \lambda_2 + 1/2)}(\zeta) \calT^{(\lambda_3 - 1/2, \, \lambda_4 + 1/2)}(\zeta) \\
- \calT^{(\lambda_1 - 1/2, \, \lambda_3 + 1/2)}(\zeta) \calT^{(\lambda_2 - 1/2, \, \lambda_4 + 1/2)}(\zeta) \\ + \calT^{(\lambda_2 - 1/2, \, \lambda_3 + 1/2)}(\zeta) \calT^{(\lambda_1 - 1/2, \, \lambda_4 + 1/2)}(\zeta) = 0, \label{utt}
\end{multline}
which we call the universal $TT$-relation.

Putting in (\ref{utt})
\begin{equation*}
\lambda_1 = \nu + 1, \qquad \lambda_2 = \nu, \qquad \lambda_3 = 0, \qquad \lambda_4 = -1,
\end{equation*}
where $\nu$ is an arbitrary complex number, we obtain
\begin{equation}
\calT^{(\nu, \, 0)}(q^{-1/s} \zeta) \, \calT^{(\nu, \, 0)}(q^{1/s}\zeta) = 1 + \calT^{(\nu - 1, \, 0)}(q^{-1/s} \zeta) \, \calT^{(\nu + 1, \, 0)}(q^{1/s} \zeta). \label{fr1}
\end{equation}
From the other hand the substitution
\begin{equation*}
\lambda_1 = \nu + 1, \qquad \lambda_2 = \nu, \qquad \lambda_3 = \nu - 1, \qquad \lambda_4 = -1,
\end{equation*}
where again $\nu$ is a complex number, gives
\begin{equation}
\calT^{(1, \, 0)}(q^{-2 \nu/s} \zeta) \, \calT^{(\nu, \, 0)}(\zeta) = \calT^{(\nu - 1, \, 0)}(\zeta) + \calT^{(\nu + 1, \, 0)}(\zeta). \label{fr2}
\end{equation}
Equations of the type (\ref{fr1}) and (\ref{fr2}) are usually called the fusion relations, see \cite{KluPea92, KunNakSuz94, BazHibKho02}.

\section{Spin chain}

\subsection{Monodromy operators}

To construct a monodromy operator we choose two representations of $\uqlslii$ for two factors of the tensor product $\uqlslii \otimes \uqlslii$. When the Jimbo's homomorphism is used for the first factor, we say that we deal with a basic monodromy operator. We restrict ourselves by using the finite dimensional representations of the form $\varphi^{(k, \, 0)}$, $k \in \bbZ_+$, and use for a general monodromy operator the notation
\begin{equation*}
M^{k, \, \ell}(\zeta | \eta) = (\id \otimes (\varphi^{(\ell, \, 0)})_\eta)(\calM^{(k, \, 0)}(\zeta)) = ((\varphi^{(k, \, 0)})_\zeta \otimes (\varphi^{(\ell, \, 0)})_\eta)(\calR).
\end{equation*}
We denote a basic monodromy operator as
\begin{equation*}
M^{\square, \, \ell}(\zeta | \eta) = (\id \otimes (\varphi^{(\ell, \, 0)})_\eta)(\calM(\zeta)) = (\varphi_\zeta \otimes (\varphi^{(\ell, \, 0)})_\eta)(\calR).
\end{equation*}
The above definitions refer to the case of a one site chain. In the case of a chain of $N$ sites we define a monodromy operator as
\begin{equation*}
M^{k, \, \ell}(\zeta | \eta_1, \ldots, \eta_N) = ((\varphi^{(k, \, 0)})_\zeta \otimes ((\varphi^{(\ell, \, 0)})_{\eta_1} \otimes_{\Delta^\op} \cdots \otimes_{\Delta^\op} (\varphi^{(\ell, \, 0)})_{\eta_N}))(\calR),
\end{equation*}
and a basic monodromy operator as
\begin{equation}
M^{\square, \, \ell}(\zeta | \eta_1, \ldots, \eta_N) = ((\varphi_\zeta \otimes ((\varphi^{(\ell, \, 0)})_{\eta_1} \otimes_{\Delta^\op} \cdots \otimes_{\Delta^\op} (\varphi^{(\ell, \, 0)})_{\eta_N}))(\calR). \label{bmo}
\end{equation}
For one site monodromy operator it follows from (\ref{ggr}) that
\begin{equation}
M^{k, \, \ell}(\zeta | \eta) = M^{k, \,  \ell}(\zeta \eta^{-1}), \qquad M^{\square, \, \ell}(\zeta | \eta) = M^{\square, \, \ell}(\zeta \eta^{-1}), \label{mzemz}
\end{equation}
where
\begin{equation*}
M^{k, \, \ell}(\zeta) = M^{k, \, \ell}(\zeta | 1), \qquad M^{\square, \, \ell}(\zeta) = M^{\square, \, \ell}(\zeta | 1).
\end{equation*}

\subsubsection{Basic monodromy operators for quantum space $(V^{(1, \, 0)})_\eta$.} \label{sss:bmo1}

In this section we consider the simplest basic monodromy operators $M^{\square, \, 1}(\zeta)$. First of all note that
\begin{equation}
\varphi_\zeta(q^{\nu h_0}) = q^{-\nu H}, \qquad \varphi_\zeta(q^{\nu h_1}) = q^{\nu H}. \label{fzh}
\end{equation}
To find the images of the root vectors under the mapping $\varphi_\zeta$ we start with the evident relations
\begin{align*}
& \varphi_\zeta(e_{\delta - \alpha}) = \zeta^{s_\delta - s_\alpha} F q^{- G_1 - G_2}, && \varphi_\zeta(e_\alpha) = \zeta^{s_\alpha} E, \\
& \varphi_\zeta(f_{\delta - \alpha}) = \zeta^{- (s_\delta - s_\alpha)} E q^{G_1 + G_2}, && \varphi_\zeta(f_\alpha) = \zeta^{- s_\alpha} F.
\end{align*}
Here and sometimes below we use the notation
\begin{equation*}
s_\delta = s_0 + s_1, \qquad s_\alpha = s_1.
\end{equation*}
Using (\ref{e2}) and (\ref{f2}), we obtain
\begin{align*}
& \varphi_\zeta(e'_\delta) = \zeta^{s_\delta} \kappa_q^{-1} q^{-1} \big[ C^{(1)} - (q + q^{-1}) (C^{(2)})^{1/2} q^{- H} \big], \\
& \varphi_\zeta(f'_\delta) = - \zeta^{-s_\delta} \kappa_q^{-1} q (C^{(2)})^{-1} \big[ C^{(1)}  - (q + q^{-1}) (C^{(2)})^{1/2} q^H \big],
\end{align*}
where $C^{(1)}$ and $C^{(2)}$ are the quantum Casimir operators defined by equations (\ref{qciia}) and (\ref{qciib}). Now equations (\ref{eapkd}) and (\ref{fapkd}) give
\begin{align}
& \varphi_\zeta(e_{\alpha + k \delta}) = \zeta^{s_\alpha + k s_\delta} (-1)^k E (C^{(2)})^{k/2} q^{- k (H + 2)}, \label{eakd} \\
& \varphi_\zeta(f_{\alpha + k \delta}) = \zeta^{- s_\alpha - k s_\delta} (-1)^k F (C^{(2)})^{-k/2} q^{k H}. \label{efkd}
\end{align}
Similarly, it follows from (\ref{edmapkd}) and (\ref{fdmapkd}) that
\begin{align}
& \varphi_\zeta(e_{(\delta - \alpha) + k \delta}) = \zeta^{(s_\delta - s_\alpha) + k s_\delta} (-1)^k F (C^{(2)})^{(k + 1)/2} q^{- k H}, \label{edmakd} \\
& \varphi_\zeta(f_{(\delta - \alpha) + k \delta}) = \zeta^{- (s_\delta - s_\alpha) - k s_\delta} (-1)^k q^{2 k} E (C^{(2)})^{-(k + 1)/2} q^{k H}. \label{fdmakd}
\end{align}
Further, equation (\ref{epkd}) gives
\begin{multline*}
\varphi_\zeta(e'_{k \delta}) = \zeta^{k s_\delta} \kappa_q^{-2} (-1)^{k - 1} q^{-k} (C^{(2)})^{(k - 1)/2} \big[(q^k - q^{-k}) \, q^{(k - 1) H} \\
- (q^{k - 1} - q^{- k + 1}) \, (C^{(2)})^{1/2} q^{- k H} - (q^{k + 1} - q^{- k - 1}) \, (C^{(2)})^{1/2} q^{-(k - 2) H} \big].
\end{multline*}
Using this equation, we determine that
\begin{multline*}
\varphi_\zeta(1 + \kappa_q e'_\delta(z)) = (1 + C^{(1)} q^{-1} \zeta^{s_\delta} z^{-1} + C^{(2)} q^{-2} \zeta^{2 s_\delta} z^{- 2}) \\*
\times (1 + (C^{(2)})^{1/2} q^{- H} \zeta^{s_\delta} z^{- 1})^{-1}(1 + (C^{(2)})^{1/2} q^{- H - 2} \zeta^{s _\delta} z^{- 1})^{-1}.
\end{multline*}
Now, it follows from the relation
\begin{equation*}
\log (1 + z^{-1}) = \sum_{k = 1}^\infty (-1)^{k - 1} \frac{z^{- k}}{k}
\end{equation*}
and from equation (\ref{edepd}) that
\begin{equation*}
\varphi_\zeta(e_{k \delta}) = \frac{\zeta^{k s_\delta}}{k} \, \kappa_q^{-1} (-1)^{k - 1} q^{- k} \big[ F_k - (q^k + q^{- k}) (C^{(2)})^{k/2} q^{- k H} \big],
\end{equation*}
where the quantities $F_k$ are determined by the generating function
\begin{equation}
\log( 1 + C^{(1)} z^{-1} + C^{(2)} z^{-2}) = \sum_{k = 1}^\infty (-1)^{k - 1} F_k \frac{z^{- k}}{k}. \label{ck2}
\end{equation}
It is clear that all $F_k$ belong to the centre of $\uqglii$. In particular, we have
\begin{equation*}
F_1 = C^{(1)}, \qquad F_2 = (C^{(1)})^2 - 2 C^{(2)}, \qquad F_3 = (C^{(1)})^3 - 3 C^{(1)} C^{(2)}.
\end{equation*}
Using equations (\ref{plc1}) and (\ref{plc2}) and (\ref{ck2}), we conclude that
\begin{equation}
\pi^\lambda(F_k) = q^{- 2(\lambda_1 + 1/2) k} + q^{- 2 (\lambda_2 - 1/2) k}. \label{plck}
\end{equation}
Finally, equation (\ref{fpkd}) gives
\begin{multline*}
\varphi_\zeta(f'_{k \delta}) = \zeta^{- k s_\delta} \kappa_q^{-2} (-1)^k q^k (C^{(2)})^{-(k + 1)/2} \big[(q^k - q^{-k}) \, C^{(1)} q^{(k - 1) H} \\*
- (q^{k - 1} - q^{- k + 1}) \, (C^{(2)})^{1/2} q^{(k - 2) H} - (q^{k + 1} - q^{- k - 1}) \, (C^{(2)})^{1/2} q^{k H} \big],
\end{multline*}
and we obtain
\begin{multline*}
\varphi_\zeta(1 - \kappa_q f'_\delta(z)) = (1 + C^{(1)} (C^{(2)})^{-1} q \zeta^{- s_\delta} z^{-1} + (C^{(2)})^{-1} q^2 \zeta^{- 2 s_\delta} z^{- 2}) \\
\times (1+ (C^{(2)})^{-1/2} q^H \zeta^{- s_\delta} z^{- 1})^{-1}(1 + (C^{(2)})^{-1/2} q^{H + 2} \zeta^{- s_\delta} z^{- 1})^{-1}.
\end{multline*}
It is not difficult to see that 
\begin{equation*}
\varphi_\zeta(f_{k \delta}) = - \frac{\zeta^{- k s_\delta}}{k} \, \kappa_q^{-1} (-1)^{k - 1} q^k \big[ F_k (C^{(2)})^{- k} - (q^k + q^{- k}) (C^{(2)})^{-k/2} q^{k H} \big].
\end{equation*}

Now we find images of the root vectors under the representation $(\varphi^{(1, \, 0)})_\eta$. Here the representation $\pi^{(1, \, 0)}$ is two dimensional. Let $\{E_{n m}\}_{n, m = 0}^1$ be the basis of the algebra $\End((V^{((1, \, 0))})_\eta)$ associated with the basis $\{v_n\}_{n = 0}^1$ of $(V^{((1, \, 0))})_\eta$. It is easy to determine that
\begin{align}
& \pi^{(1, \, 0)}(q^{\nu G_1}) = q^{\nu} E_{00} + E_{11}, && \pi^{(1, \, 0)}(q^{\nu G_2}) = E_{00} + q^{\nu} E_{11}, \label{qh1} \\
& \pi^{(1, \, 0)}(E) = E_{01}, && \pi^{(1, \, 0)}(F) = E_{10}. \label{ef1}
\end{align}
It follows from (\ref{fzh}) that
\begin{equation}
(\varphi^{(1, \, 0)})_\eta(q^{\nu h_0}) = q^{-\nu} E_{00} + q^\nu E_{11}, \qquad (\varphi^{(1, \, 0)})_\eta(q^{\nu h_1}) = q^\nu E_{00} + q^{- \nu} E_{11}. \label{phi10qh}
\end{equation}
Using equations (\ref{eakd})--(\ref{fdmakd}), one can demonstrate that
\begin{align*}
(\varphi^{(1, \, 0)})_\eta(e_{\alpha + k \delta}) & = \eta^{s_\alpha + k s_\delta} (-1)^k q^{- 2 k} E_{01}, \\
(\varphi^{(1, \, 0)})_\eta(e_{(\delta - \alpha) + k \delta}) & = \eta^{(s_\delta - s_\alpha) + k s_\delta} (-1)^k q^{- 2 k - 1} E_{10}, \\
(\varphi^{(1, \, 0)})_\eta(f_{\alpha + k \delta}) & = \eta^{- s_\alpha - k s_\delta} (-1)^k q^{2 k} E_{10}, \\
(\varphi^{(1, \, 0)})_\eta(f_{(\delta - \alpha) + k \delta}) & = \eta^{- (s_\delta - s_\alpha) + k s_\delta} (-1)^k q^{2 k + 1} E_{01}.
\end{align*}
Equation (\ref{plck}) takes the form
\begin{equation*}
\pi^{(1, \, 0)}(F_k) = q^{- k}(q^{2 k} + q^{-2 k}), 
\end{equation*}
and we have
\begin{align*}
(\varphi^{(1, \, 0)})_\eta(e_{k \delta}) & = \frac{\eta^{k s_\delta}}{k} \, (-1)^{k - 1} q^{- k} [k]_q (E_{00} - q^{-2 k} E_{11}), \\
(\varphi^{(1, \, 0)})_\eta(f_{k \delta}) & = \frac{\eta^{- k s_\delta}}{k} \, (-1)^{k - 1} q^k [k]_q (E_{00} - q^{2 k} E_{11}).
\end{align*}
Using the expression for the universal $R$-matrix given in section \ref{sss:urm}, the above relations for the images of the root vectors of $\uqlslii$, and the equation
\begin{equation*}
\calK_{\varphi_\zeta, \, (\varphi^{(1, \, 0))})_\eta} = q^{(G_1 - G_2)/2} E_{00} +  q^{-(G_1 - G_2)/2} E_{11},
\end{equation*}
we obtain
\begin{multline}
M^{\square, \, 1}(\zeta) = \exp ({F_2(\zeta^{s_\delta})}) (C^{(2)})^{1/4} \Big[
q^{G_1} (1 - \zeta^{s_\delta} q^{- 2 G_1}) \otimes E_{00} \\+ \zeta^{s_\delta - s_\alpha} \kappa_q q^{- G_1} F \otimes E_{01} + \zeta^{s_\alpha} \kappa_q E q^{G_1} \otimes E_{10} + q^{G_2} (1 - \zeta^{s_\delta} q^{- 2 G_2}) \otimes E_{11} \Big], \label{bmo1}
\end{multline}
where
\begin{equation*}
F_2(\zeta) = \sum_{k = 1}^\infty \frac{1}{q^k + q^{-k}} \, F_k \frac{\zeta^k}{k}.
\end{equation*}
It is useful to have in mind that
\begin{equation*}
F_2(q \, \zeta) + F_2(q^{-1} \zeta) = - \log(1 - C^{(1)} \zeta + C^{(2)} \zeta^2).
\end{equation*}
It also follows from (\ref{plck}) that
\begin{equation*}
\pi^\lambda(F_2(\zeta)) = f_2(q^{- 2(\lambda_1 + 1/2)} \zeta) + f_2(q^{- 2(\lambda_2 - 1/2)} \zeta),
\end{equation*}
where $f_2(\zeta)$ is the transcendental function defined as
\begin{equation*}
f_2(\zeta) = \sum_{k = 1}^\infty \frac{1}{q^k + q^{-k}} \, \frac{\zeta^k}{k}.
\end{equation*}
This function satisfies the following defining equation
\begin{equation*}
f_2(q \zeta) + f_2(q^{-1} \zeta) = - \log(1 - \zeta)
\end{equation*}
with the initial condition $f_2(0) = 0$.

For any non-negative integer $\ell$ we represent the corresponding basic monodromy operator as
\begin{equation}
M^{\square, \, \ell}(\zeta) = \sum_{n, m} \bbM^{\square, \, \ell}(\zeta)_{n m} \otimes E_{n m} \label{mbbm}
\end{equation}
and denote the matrix formed by $\bbM^{\square, \, \ell}(\zeta)_{n m}$ by $\bbM^{\square, \, \ell}(\zeta)$. Note that $\bbM^{\square, \, \ell}(\zeta)_{n m}$ are elements of $\uqglii$. For the case of $\ell = 1$ we have
\begin{equation*}
\bbM^{\square, \, 1}(\zeta) = \exp ({F_2(\zeta^{s_\delta})}) (C^{(2)})^{1/4} \left( \begin{array}{cc}
q^{G_1} (1 - \zeta^{s_\delta} q^{- 2 G_1}) & \zeta^{s_\delta - s_\alpha} \kappa_q q^{- G_1} F \\[.5em]
\zeta^{s_\alpha} \kappa_q E q^{G_1} & q^{G_2} (1 - \zeta^{s_\delta} q^{- 2 G_2}) 
\end{array} \right).
\end{equation*}

\subsubsection{Basic monodromy operators for quantum space $(V^{(\ell, \, 0))_\eta}$.}

One can construct the monodromy operator for the quantum space $(V^{(\ell, \, 0))_\eta}$ for $\ell > 1$ in the same way as it is done for the case of the quantum space $(V^{(1, 0)})_\eta$ in the preceding section . However, there is a simpler recursive way usually called fusion \cite{KulResSkl81, KulSkl82b, Jim89}.

Let us show that the representation $(\varphi^{(\ell - 1, \, 0)})_{\eta_1} \otimes_{\Delta^\op} (\varphi^{(1, \, 0)})_{\eta_2}$, with an appropriate choice of $\eta_1$, $\eta_2$ and $\eta$, has a subrepresentation isomorphic to the representation $(\varphi^{(\ell, \, 0)})_\eta$. To this end first observe that the basis $\{v_n\}$ of the module $(V^{(\ell, \, 0)})_\eta$ can be constructed as follows. We start with a vector $v_0$ satisfying the conditions
\begin{equation}
q^{\nu h_0} v_0 = q^{- \nu \ell} v_0, \qquad q^{\nu h_1} v_0 = q^{\nu \ell} v_0, \qquad e_1 v_0 = 0, \label{dv0}
\end{equation}
and define
\begin{equation*}
v_1 = \eta^{-s_0} q^\ell \, e_0 \, v_0, \qquad \ldots, \qquad v_\ell = \eta^{-s_0} q^\ell \, e_0 \, v_{\ell - 1}.
\end{equation*}

The module $(V^{(\ell - 1, \, 0)})_{\eta_1} \otimes_{\Delta^\op} (V^{(1, \, 0)})_{\eta_2}$ contains the vector $v_0 \otimes v_0$ which satisfies the same conditions (\ref{dv0}) as the vector $v_0$. Denote this vector by $w_0(\eta)$ and put 
\begin{equation*}
w_1(\eta) = \eta^{-s_0} q^\ell \, e_0 \, w_0(\eta), \qquad \ldots, \qquad w_\ell(\eta) = \eta^{-s_0} q^\ell \, e_0 \, w_{\ell - 1}(\eta).
\end{equation*}
The explicit expression for the vectors $w_n$, $n = 0, 1, \ldots, \ell$, is
\begin{equation*}
w_n(\eta) = \eta^{- n s_0} \eta_1^{(n - 1) s_0} \eta_2^{s_0} q^{n + \ell - 2} [n]_q \, v_{n - 1} \otimes v_1 + \eta^{- n s_0} \eta_1^{n s_0} q^{2 n} \, v_n \otimes v_0,
\end{equation*}
where we assume that $v_{\ell + 1} \otimes v_0 = 0$.

For arbitrary values of $\eta_1$ and $\eta_2$ the linear span of the vectors $w_0$, $w_1$, $\ldots$, $w_\ell$ is not invariant under the action of the generators of $\uqlslii$. In particular, we have
\begin{equation*}
e_1 w_\ell(\eta) = \eta^{-2 \ell s_0} \eta_1^{(\ell - 1) s_0} \eta_2^{s_0} q^{2 (\ell - 1)} [\ell]_q \big( \eta_1^{s_1} q [\ell - 1]_q \, v_{\ell - 2} \otimes v_1 + \eta_2^{s_1} \, v_{\ell - 1} \otimes v_0 \big).
\end{equation*}
Hence, $e_1 w_\ell(\eta)$ is not proportional to
\begin{equation*}
w_{\ell - 1}(\eta) = \eta^{- (\ell - 1) s_0} \eta_1^{- (\ell - 2) s_0} q^{2 \ell - 2} \big(  \eta_2^{s_0} q^{-1} [\ell - 1]_q \, v_{\ell - 2} \otimes v_1 + \eta_1^{s_0} \, v_{\ell - 1} \otimes v_0 \big),
\end{equation*}
as it should be. However, assuming that
\begin{equation*}
\eta_1 = q^{-2/s} \eta, \qquad \eta_2 = \eta, 
\end{equation*}
we obtain
\begin{equation*}
e_1 w_\ell(\eta) = \eta^{s_1} [\ell]_q \, w_{\ell - 1}(\eta).
\end{equation*}
One can easily see that in this case the vectors $w_n(\eta)$ do not depend on $\eta$, so we write below just $w_n$. One can demonstrate that the vectors $w_0$, $w_1$, $\ldots$, $w_\ell$ form a basis of the submodule of $(V^{(\ell - 1, \, 0)})_{q^{-2/s} \eta} \otimes_{\Delta^\op} (V^{(1, \, 0)})_{\eta}$ equivalent to the basis of the module $(V^{(\ell, \, 0)})_\eta$ formed by the vectors $v_0$, $v_1$, $\ldots$, $v_\ell$.

Denote by $S$ the linear mapping from $(V^{(\ell, \, 0)})_\eta$ to $(V^{(\ell - 1, \, 0)})_{q^{-2/s} \eta} \otimes_{\Delta^\op} (V^{(1, \, 0)})_{\eta}$ defined by the relation
\begin{equation*}
S(v_n) = w_n,
\end{equation*}
and by $S'$ the left inverse of $S$,
\begin{equation*}
S' \circ S = \id.
\end{equation*}
Notice that the mapping $S'$ is not unique. However, one can use any mapping $S'$ satisfying the above equation. It is clear that for any choice we obtain
\begin{equation*}
S'(w_n) = v_n,
\end{equation*}
and have
\begin{equation*}
S' \circ \big(((\varphi^{(1, \, 0)})_{q^{-2/s} \eta} \otimes_{\Delta^\op} (\varphi^{(1, \, 0)})_{\eta})(a)\big) \circ S = (\varphi^{(2, \, 0)})_\eta(a)
\end{equation*}
for any $a \in \uqlslii$. Using this relation, we obtain
\begin{multline}
M^{\square, \, \ell}(\zeta | \eta) = (\varphi_\zeta \otimes (\varphi^{(\ell, \, 0)})_\eta)(\calR) \\
= (\id \otimes F_{S', \, S}) \big((\varphi_\zeta \otimes ((\varphi^{(\ell - 1, \, 0)})_{q^{-2/s} \eta} \otimes_{\Delta^\op} (\varphi^{(1, \, 0)})_{\eta}))(\calR)\big), \label{msl}
\end{multline}
where $F_{S', S}$ is a mapping from $\End((V^{(\ell - 1, \, 0)})_{q^{-2/s} \eta} \otimes_{\Delta^\op} (V^{(1, \, 0)})_{\eta})$ to $\End(V^{(\ell, \, 0)})_\eta$ defined as
\begin{equation*}
F_{S', \, S}(A) = S' \circ A \circ S.
\end{equation*}
Apply the mapping $\id \otimes \Pi^{23}$ to both sides of the second equation of (\ref{urm}). This gives
\begin{equation*}
(\id \otimes \Delta^\op)(\calR) = (\id \otimes \Pi^{23}) (\calR^{13} \calR^{12}) = \calR^{12} \calR^{13}.
\end{equation*}
Now, it follows from (\ref{msl}) that 
\begin{equation*}
M^{\square, \, \ell}(\zeta | \eta) = (\id \otimes F_{S', \, S}) \big( (M^{\square, \, \ell - 1})^{12}(\zeta | q^{-2/s} \eta)  (M^{\square, \, 1})^{13}(\zeta | \eta) \big).
\end{equation*}
Having in mind the representation (\ref{mbbm}), we rewrite this equation as
\begin{multline*}
M^{\square, \, \ell}(\zeta | \eta) \\= \sum_{n_1, n_2, m_1, m_2} \bbM^{\square, \, \ell - 1}(\zeta | q^{-2/s} \eta)_{n_1 m_1} \bbM^{\square, \, 1}(\zeta | q^{-2/s} \eta)_{n_2 m_2} \otimes (S' \circ (E_{n_1 m_1} \otimes E_{n_2 m_2}) \circ S).
\end{multline*}
It is not difficult to determine that
\begin{equation*}
S' \circ (E_{n_1 m_1} \otimes E_{n_2 m_2}) \circ S = \sum_{n, m} \bbS'_{n | n_1 n_2} \bbS_{m_1 m_2 | m} \, E_{n m},
\end{equation*}
where the quantities $\bbS'_{n | n_1 n_2}$ and $\bbS_{m_1 m_2 | m}$ are defined by the equations
\begin{equation*}
S'(v_{n_1} \otimes v_{n_2}) = \sum_n v_n \, \bbS'_{n | n_1 n_2}, \qquad S(w_m) = \sum_{m_1, m_2} v_{m_1} \otimes v_{m_2} \, \bbS_{m_1 m_2 | m}.
\end{equation*}
Finally, we come to the relation 
\begin{equation*}
\bbM^{\square, \, \ell}(\zeta) = \bbS' \, (\bbM^{\square, \, \ell - 1}(q^{2/s} \zeta) \otimes \bbM^{\square, \, 1}(\zeta)) \, \bbS,
\end{equation*}
where $\bbS'$ and $\bbS$ are the matrices formed by the matrix entries $\bbS'_{n | n_1 n_2}$ and $\bbS_{m_1 m_2 | m}$. The above equation shows that we can recursively construct $\bbM^{\square, \, \ell}(\zeta | \eta)$ from $\bbM^{\square, \, 1}(\zeta | \eta)$. Notice that the matrix $\bbS$ is unique, while there is some freedom in the choice of the matrix $\bbS'$. For the case of $\ell = 2$ one can use
\begin{equation}
\bbS' = \left( \begin{array}{cccc}
1 & 0 & 0 & 0\\
0 & q^{-1} & 0 & 0\\
0 & 0 & 0 & q^{- 2 s_1/s} [2]_q^{-1}
\end{array} \right)
\qquad
\bbS = \left( \begin{array}{ccc}
1 & 0 & 0 \\
0 & q & 0 \\
0 & q^{2 s_1/s} & 0 \\
0 & 0 & q^{2 s_1/s} [2]_q
\end{array} \right). \label{sps}
\end{equation}
For the case of $\ell = 0$ one comes to the following expression for the basic monodromy matrix
\begin{align}
M^{\square, \, 2}&(\zeta) = (1 - C^{(1)} q \zeta^s + C^{(2)} q^2 \zeta^{2 s})^{-1} (C^{(2)})^{1/2} \notag \\
\times \Big[ & q^{2 G_1}(1 - q^{- 2 G_1} \zeta^s)(1 - q^2 q^{- 2 G_1} \zeta^s) E_{00} + \kappa_q^2 q^2 F (1 - q^2 q^{- 2 G_1} \zeta^s) \zeta^{s_0} E_{01} \notag \\
& + \kappa_q^2 q^5 [2]_q F^2 q^{- 2 G_1} \zeta^{2 s_0} E_{02} + \kappa_q E (1 - q^{- 2 G_1} \zeta^s) q^{2 G_1} \zeta^{s_1} E_{10} \notag \\
& + q^{G_1 + G_2}(1 + q(C^{(1)} - [2]_q q^{- 2 G_1} - [2]_q q^{- 2 G_2})\zeta^s + q^2 q^{- 2 (G_1 + G_2)} \zeta^{2s}) E_{11} \notag \\
& + \kappa_q q^2 F (1 - q^{- 2 G_2} \zeta^s) q^{- G_1 + G_2} \zeta^{s_0} E_{12} + \kappa_q^2 q [2]_q^{-1} E^2 q^{2 G_1} \zeta^{2 s_1} E_{20} \notag \\
& + \kappa_q E (1 - q^2 q^{- 2 G_2} \zeta^s) q^{G_1 + G_2} \zeta^{s_1} E_{21} + q^{2 G_2}(1 - q^{- 2 G_2} \zeta^s)(1 - q^2 q^{- 2 G_2} \zeta^s) E_{22} \Big]. \label{bmo2}
\end{align}

\subsection{\texorpdfstring{$R$-operators}{R-operators}}

Usually a monodromy operator of the type $M^{\ell, \, \ell}(\zeta)$ is called an $R$-operator and denoted as $R^\ell(\zeta)$. The $R$-operators play a special role in the study of integrable systems. The $R$-operator $R^\ell(\zeta)$ is obtained from the basic monodromy operator $M^{\square, \, \ell}(\zeta)$ by applying the corresponding representation $\pi^{(\ell, \, 0)}$.

For the case of $\ell = 1$, using the expression (\ref{bmo1}) for the basic monodromy operator $M^{\square, \, 1}(\zeta)$ and equations (\ref{qh1}) and (\ref{ef1}) describing the representation $\pi^{(1, \, 0)}$, we obtain the well known result
\begin{multline*}
R^1(\zeta) = M^{1, \, 1}(\zeta) = \exp(f_2(q^{-3} \zeta^s) + f_2(q \zeta)) q^{-1/2} \\
\times \Big[ q (1 - q^{-2} \zeta^s) (E_{00} \otimes E_{00} + E_{11} \otimes E_{11}) + (1 - \zeta^s) (E_{00} \otimes E_{11} + E_{11} E_{00}) \\
+ \kappa_q (\zeta^{s_1} E_{01} \otimes E_{10} + \zeta^{s_0} E_{10} \otimes E_{01}) \Big].
\end{multline*}
The representation $\pi^{(2, \, 0)}$ is described by the equations 
\begin{align*}
& \pi^{(2, \, 0)}(q^{\nu G_1}) = q^{2 \nu} E_{00} + q E_{11} + E_{22}, && \pi^{(2, \, 0)}(q^{\nu G_2}) = E_{00} + q^{\nu} E_{11} + q^{2 \nu} E_{22}, \\
& \pi^{(2, \, 0)}(E) = [2]_q E_{01} + [2]_q E_{12}, && \pi^{(2, \, 0)}(F) = E_{10} + E_{21},
\end{align*}
and, using (\ref{bmo2}), we obtain
\begin{align*}
R^2(\zeta)& = M^{2, \, 2}(\zeta) = (1 - q^2 \zeta^s)^{- 1} (1 - q^{- 4} \zeta^s)^{- 1} q^{-2} \\*
\times \Big[ & q^4 (1 - q^{-2} \zeta^s)(1 - q^{-4} \zeta^s) (E_{00} \otimes E_{00} + E_{22} \otimes E_{22}) \\
& + q^2 (1 - \zeta^s)(1 - q^{-2} \zeta^s) (E_{00} \otimes E_{11} + E_{11} \otimes E_{00} + E_{11} \otimes E_{22} + E_{22} \otimes E_{11}) \\
& + (1 - \zeta^s)(1 - q^2 \zeta^s)(E_{00} \otimes E_{22} + E_{22} \otimes E_{00}) \\
& + \kappa_q [2]_q q^2 (1 - q^{-2} \zeta^s) (\zeta^{s_1} E_{01} \otimes E_{10} + \zeta^{s_0} E_{10} \otimes E_{01} + \zeta^{s_1} E_{12} \otimes E_{21} + \zeta^{s_0} E_{21} \otimes E_{12}) \\
& + \kappa_q [2]_q (1 - \zeta^s) (\zeta^{s_0} q^2 E_{01} \otimes E_{21} + \zeta^{s_1} E_{10} \otimes E_{12} + \zeta^{s_0} E_{12} \otimes E_{10} + \zeta^{s_1} q^2 E_{21} \otimes E_{01}) \\
& + \kappa_q^2 q [2]_q (\zeta^{2 s_1} E_{02} \otimes E_{20} + \zeta^{2 s_0} E_{20} \otimes E_{02}) \\
& + q^{-2}(q^4 + (1 - 2 q^2 - 2 q^4 + q^6) \zeta^s + q^2 \zeta^{2 s}) E_{11} \otimes E_{11} \Big].
\end{align*}
This expression is consistent with the $R$-matrix of the Zamolodchikov--Fateev model
\cite{ZamFat80}.
 
\subsection{\texorpdfstring{$L$-operators}{L-operators}}

The $L$-operators corresponding to the quantum space $(V^{(\ell, \, 0)})_\eta$ are defined as
\begin{equation*}
L'^\ell_i(\zeta | \eta) = (\id \otimes (\varphi^{(\ell, \, 0)})_\eta)(\calL'_i(\zeta)) = ((\rho_i)_\zeta \otimes (\varphi^{(\ell, \, 0)})_\eta)(\calR)
\end{equation*}
for one site chain, and as
\begin{multline}
L'^\ell_i(\zeta | \eta_1, \ldots, \eta_N) = (\id \otimes ((\varphi^{(\ell, \, 0)})_{\eta_1} \otimes_{\Delta^\op} \cdots \otimes_{\Delta^\op} (\varphi^{(\ell, \, 0)})_{\eta_N})(\calL'_i(\zeta)) \\= ((\rho_i)_\zeta \otimes ((\varphi^{(\ell, \, 0)})_{\eta_1} \otimes_{\Delta^\op} \cdots \otimes_{\Delta^\op} (\varphi^{(\ell, \, 0)})_{\eta_N})(\calR) \label{lpli}
\end{multline}
for a chain of $N$ sites. As for the case of the basic one site monodromy operators we have
\begin{equation*}
L'^\ell_i(\zeta | \eta) = L'^\ell_i(\zeta \eta^{-1})
\end{equation*}
where $L'^\ell_i(\zeta) = L'^\ell_i(\zeta | 1)$.

Construct the $L$-operators for the case of the quantum space $(V^{(1, \, 0)})_\eta$. First observe that for any $a \in \uqlslii$ we have
\begin{equation*}
((\varphi^{(1, \, 0)})_\eta \circ \sigma)(a) = O \, (\varphi^{(1, \, 0)})_\eta(a) \, O^{-1}|_{s \to \sigma(s)},
\end{equation*}
where $O$ is a linear operator on $(V^{(1, \, 0)})_\eta$ having the matrix form
\begin{equation*}
\bbO = \left( \begin{array}{cc}
0 & q \\
1 & 0
\end{array} \right).
\end{equation*}
Using equation (\ref{lipo}), we see now that
\begin{equation}
\bbL'^1_{i + 1}(\zeta) = \bbO \, \bbL'^1_i(\zeta) \, \bbO^{-1}|_{s \to \sigma(s)}. \label{bblipo}
\end{equation}

It is not difficult to find the necessary images of root vectors under $\rho_\zeta$. Indeed, start with the equations
\begin{equation*}
\rho_\zeta(e_\alpha) = \zeta^{s_\alpha} \kappa_q^{-1} b \, q^{- N}, \qquad  \rho_\zeta(e_{\delta - \alpha}) = \zeta^{s_\delta - s_\alpha} b^\dagger.
\end{equation*}
Then one easily determines that definition (\ref{e2}) together with (\ref{rhoe}) give
\begin{equation}
\rho_\zeta(e'_\delta) = \kappa_q^{-1} q^{-1} \zeta^s,
\label{cepd}
\end{equation}
and, using (\ref{eapkd}) and (\ref{edmapkd}), we immediately obtain
\begin{equation}
\rho_\zeta(e_{\alpha + k \delta}) = 0, \qquad
\rho_\zeta(e_{(\delta - \alpha) + k \delta}) = 0, \qquad k \ge 1.
\label{ceamd1}
\end{equation}
The definition (\ref{epkd}) and equations (\ref{ceamd1}) give
\begin{equation*}
\rho_\zeta(e'_{k \delta}) = 0, \qquad k > 1,
\end{equation*}
and one finds that
\begin{equation*}
\rho_\zeta(e_{k \delta}) = (-1)^{k - 1} \kappa_q^{-1} q^{-k} \, \frac{\zeta^{k s}}{k}.
\end{equation*}
The necessary images of root vectors under $\varphi^{(1, \, 0)}_\eta$ are given in section \ref{sss:bmo1}. After all, having in mind (\ref{kpipii}), (\ref{phi10qh}) and (\ref{rhoh}), we observe that
\begin{equation}
\calK_{\rho_\zeta, \, (\varphi^{(1, \, 0)})_\eta} = q^{- N} \otimes E_{00} + q^N \otimes E_{11}.
\label{ok1}
\end{equation}
All that allows us to find the expression for the $L$-operator $L'^1_2(\zeta)$. To obtain the expression for $L'^1_1(\zeta)$ one can use equation (\ref{bblipo}).

The explicit form of the $L$-operators $L'^1_1(\zeta)$ and $L'^1_s(\zeta)$ is
\begin{align}
& L'^1_1(\zeta) = \rme^{f_2(\zeta^s)} [(q^N - q^{- N - 1} \zeta^s) \otimes E_{00} + b \, q^{- 2 N + 1} \zeta^{s_0} \otimes E_{01} \notag \\*
& \hspace{11.em} {} +  \kappa_q \,  b^\dagger \, q^N \zeta^{s_1} \otimes E_{10} + q^{- N} \otimes E_{11}], \label{lp11} \\
& L'^1_2(\zeta) = \rme^{f_2(\zeta^s)} [q^{- N} \otimes E_{00} + \kappa_q \,  b^\dagger \, q^{N + 1} \zeta^{s_0} \otimes E_{01} \notag \\*
& \hspace{13.5em} {} + b \, q^{- 2 N} \zeta^{s_1} \otimes E_{10} + (q^N - q^{- N - 1} \zeta^s) \otimes E_{11}]. \label{lp21}
\end{align}
The corresponding matrices look as
\begin{align*}
& \bbL'^1_1(\zeta) = \rme^{f_2(\zeta^s)} \left( \begin{array}{cc}
 q^N - q^{- N - 1} \zeta^s & b \, q^{- 2 N + 1} \zeta^{s_0} \\[.5em]
\kappa_q \,  b^\dagger \, q^N \zeta^{s_1} & q^{- N}
\end{array} \right), \\[.5em]
& \bbL'^1_2(\zeta) = \rme^{f_2(\zeta^s)} \left( \begin{array}{cc}
q^{- N} & \kappa_q \,  b^\dagger \, q^{N + 1} \zeta^{s_0} \\[.5em]
b \, q^{- 2 N} \zeta^{s_1} & q^N - q^{- N - 1} \zeta^s
\end{array} \right).
\end{align*}

To obtain the $L$-operators for the quantum space $(V^{(\ell, 0)})_\eta$ with $\ell > 1$ one can use the same fusion procedure as for the basic monodromy operators. The main relation here is very similar,
\begin{equation*}
\bbL'^\ell_i(\zeta) = \bbS' \, (\bbL'^{\ell - 1}_i(q^{2/s} \zeta) \otimes \bbL'^1_i(\zeta)) \, \bbS
\end{equation*}
with $\bbS'$ and $\bbS$ again given by equation (\ref{sps}).

Now, using equation (\ref{lp11}), we obtain the expression
\begin{align*}
L^2_1(\zeta) = (1 & - q \zeta^s)^{-1} \Big[ q^{2 N} (1 - q q^{- 2 N} \zeta^s) (1 - q^{- 1} q^{- 2 N} \zeta^s) \otimes E_{00} \\
& + q^2 [2]_q b (1 - q q^{- 2 N} \zeta^s) q^{- N} \zeta^{s_0} \otimes E_{01} + q^6 [2]_q  b^2 q^{- 4 N} \zeta^{2 s_0} \otimes E_{02} \\
& + \kappa_q b^\dagger (1 - q^{-1} q^{- 2 N} \zeta^s) q^{2 N} \zeta^{s_1} \otimes E_{10} + (1 + (q - [2]_q q^{- 2 N}) \zeta^s) \otimes E_{11} \\
& + q^2 [2]_q b q^{- 3 N} \zeta^{s_0} \otimes E_{12} + \kappa_q^2 q [2]_q^{-1} (b^\dagger)^2 q^{2 N} \zeta^{2 s_1} \otimes E_{20} \\
& \hspace{17.em} {} + \kappa_q b^\dagger \zeta^{s_1} \otimes E_{21} + q^{- 2 N} \otimes E_{22} \Big],
\end{align*}
and, using (\ref{lp21}), the expression
\begin{align*}
L^2_2(\zeta) & = (1 - q \zeta^s)^{-1} \Big[ q^{- 2 N} \otimes E_{00} + \kappa_q q^2 [2]_q b^\dagger \zeta^{s_0} \otimes E_{01} \\
& + \kappa_q^2 q^5 [2]_q (b^\dagger)^2 q^{2 N} \zeta^{2 s_0} \otimes E_{02} + b q^{- 3 N} \zeta^{s_1} \otimes E_{10} \\
& + (1 + (q - [2]_q q^{- 2 N}) \zeta^s) \otimes E_{11} + \kappa_q q^2 [2]_q b^\dagger (1 - q^{-1} q^{- 2 N} \zeta^s) q^{2 N} \zeta^{s_0} \otimes E_{12} \\
& + q^2 [2]_q^{-1} b^2 q^{- 4 N} \zeta^{2 s_1} \otimes E_{20} + b(1 - q q^{- 2 N} \zeta^s) q^{- N} \zeta^{s_1} \otimes E_{21} \\
& \hspace{13.em} {} +  q^{2 N} (1 - q q^{- 2 N} \zeta^s) (1 - q^{- 1} q^{- 2 N} \zeta^s) \otimes E_{22} \Big].
\end{align*}

\subsection{\texorpdfstring{Transfer operators and $Q$-operators}{Transfer operators and Q-operators}}

We define the transfer operator corresponding to the representations $\pi^{(k, \, 0)}$ and $\pi^{(\ell, \, 0)}$ of $\uqglii$ by the equation
\begin{multline*}
T^{k, \ell}(\zeta | \eta_1, \ldots, \eta_N) = ((\varphi^{(\ell, \, 0)})_{\eta_1} \otimes_{\Delta^\op} \cdots \otimes_{\Delta^\op} (\varphi^{(\ell, \, 0)})_{\eta_N})(\calT^{(k, \, 0)}(\zeta)) \\
= (\tr^{(k, \, 0)} \otimes ((\varphi^{(\ell, \, 0)})_{\eta_1} \otimes_{\Delta^\op} \cdots \otimes_{\Delta^\op} (\varphi^{(\ell, \, 0)})_{\eta_N}))(\calM(\zeta)).
\end{multline*}
Using the definition (\ref{bmo}) of the basic monodromy operator $M^{\square, \, \ell}(\zeta | \eta_1, \ldots, \eta_N)$, we can write
\begin{equation*}
T^{k, \, \ell}(\zeta | \eta_1, \ldots, \eta_N) = (\tr^{(k, \, 0)} \otimes \id)(M^{\square, \, \ell}(\zeta | \eta_1, \ldots, \eta_N) (\varphi_\zeta(t) \otimes 1)).
\end{equation*}
Represent the transfer operator $T^{k, \, \ell}(\zeta | \eta_1, \ldots, \eta_N)$ as
\begin{equation*}
T^{k, \, \ell}(\zeta | \eta_1, \ldots, \eta_N) = \sum \bbT^{k, \, \ell}(\zeta | \eta_1, \ldots, \eta_N)_{n_1 \ldots n_N | m_1 \ldots m_N} \, E_{n_1 m_1} \otimes \cdots \otimes E_{n_N m_N}
\end{equation*}
and define the transfer matrix
\begin{equation*}
\bbT^{k, \, \ell}(\zeta | \eta_1, \ldots, \eta_N) = (\bbT^{k, \, \ell}(\zeta | \eta_1, \ldots, \eta_N)_{n_1 \ldots n_N | m_1 \ldots m_N}).
\end{equation*}
Now, using the equality
\begin{equation*}
M^{\square, \, \ell}(\zeta | \eta_1, \ldots, \eta_N) = (M^{\square, \, \ell})^{01}(\zeta | \eta_1) \ldots (M^{\square, \, \ell})^{0N}(\zeta | \eta_N),
\end{equation*}
one can demonstrate that
\begin{equation*}
\bbT^{k, \, \ell}(\zeta | \eta_1, \ldots, \eta_N) = \tr^{(k, \, 0)}((\bbM^{\square, \, \ell}(\zeta \eta_1^{-1}) \otimes \cdots \otimes \bbM^{\square, \, \ell}(\zeta \eta_N^{-1})) \, q^{ - \Phi_1 G_1 - \Phi_2 G_2}),
\end{equation*}
see, for example, the paper \cite{BooGoeKluNirRaz14a}. Here we use the fact that
\begin{equation*}
\varphi(t) = q^{ - \Phi_1 G_1 - \Phi_2 G_2},
\end{equation*}
where we denote
\begin{equation*}
\Phi_1 = (\phi_0 - \phi_1)/4, \qquad \Phi_2 = (\phi_1 - \phi_0)/4.
\end{equation*}

We define the $Q$-operators corresponding to the representation $\pi^{(\ell, \, 0)}$ of $\uqglii$ as
\begin{multline*}
Q^\ell_i(\zeta | \eta_1, \ldots, \eta_N) = ((\varphi^{(\ell, \, 0)})_{\eta_1} \otimes_{\Delta^\op} \cdots \otimes_{\Delta^\op} (\varphi^{(\ell, \, 0)})_{\eta_N})(\calQ_i(\zeta)) \\= ((\varphi^{(\ell, \, 0)})_{\eta_1} \otimes_{\Delta^\op} \cdots \otimes_{\Delta^\op} (\varphi^{(\ell, \, 0)})_{\eta_N})(\calQ'_i(\zeta) \zeta^{\calD_i}) \\
= (\tr^{\scriptscriptstyle +} \otimes ((\varphi^{(\ell, \, 0)})_{\eta_1} \otimes_{\Delta^\op} \cdots \otimes_{\Delta^\op} (\varphi^{(\ell, \, 0)})_{\eta_N}))(\calL'_i(\zeta)((\rho_i)_\zeta(t) \otimes \zeta^{\calD_i})).
\end{multline*}
It is easy to understand that
\begin{equation*}
((\varphi^{(\ell, \, 0)})_{\eta_1} \otimes_{\Delta^\op} \cdots \otimes_{\Delta^\op} (\varphi^{(\ell, \, 0)})_{\eta_N})(\zeta^{\calD_i}) = q^{- (N - 1) \Phi_i s /2} \, \zeta^{D_i^\ell} \otimes \cdots \otimes \zeta^{D_i^\ell},
\end{equation*}
where $\zeta^{D_i^\ell}$ is an element of $\End((V^{(\ell, \, 0)})_\eta)$ defined by the equation
\begin{equation*}
(\varphi^{(1, \, 0)})_\eta(\zeta^{\calD_i}) = \zeta^{D_i^\ell}.
\end{equation*}
Hence, using (\ref{lpli}), we come to the equation 
\begin{multline}
Q^\ell_i(\zeta | \eta_1, \ldots, \eta_N) \\= q^{- (N - 1) \Phi_i s /2} (\tr^{\scriptscriptstyle +} \otimes \id)(L'_i(\zeta | \eta_1, \ldots, \eta_N) ((\rho_i)_\zeta(t) \otimes (\zeta^{D_i^\ell} \otimes \cdots \otimes \zeta^{D_i^\ell}))). \label{qilns}
\end{multline}

Represent the $Q$-operator $Q^\ell_i(\zeta | \eta_1, \ldots, \eta_N)$ as
\begin{equation*}
Q^\ell_i(\zeta | \eta_1, \ldots, \eta_N) = \sum \bbQ^\ell_i(\zeta | \eta_1, \ldots, \eta_N)_{n_1 \ldots n_N | m_1 \ldots m_N} \, E_{n_1 m_1} \otimes \cdots \otimes E_{n_N m_N},
\end{equation*}
and define the matrix
\begin{equation*}
\bbQ^\ell_i(\zeta | \eta_1, \ldots, \eta_N) = (\bbQ^\ell_i(\zeta | \eta_1, \ldots, \eta_N)_{n_1 \ldots n_N | m_1 \ldots m_N}).
\end{equation*}
One can get convinced that 
\begin{equation*}
L'^\ell_i(\zeta | \eta_1, \ldots, \eta_N) = ((L'^\ell_i)^{01}(\zeta | \eta_1) \ldots (L'^\ell_i)^{0N}(\zeta | \eta_N)),
\end{equation*}
see, for example, the paper \cite{BooGoeKluNirRaz14a}. After all, using (\ref{qilns}), we come to the equation
\begin{equation*}
\bbQ^\ell_i(\zeta | \eta_1, \ldots, \eta_N) = q^{- (n - 1) \Phi_i s /2} \tr^{\scriptscriptstyle +}((\bbL'^\ell_i(\zeta \eta_1^{-1}) \zeta^{\bbD_i^\ell} \otimes \cdots \otimes \bbL'^\ell_i(\zeta \eta_N^{-1}) \zeta^{\bbD_i^\ell}) \, (\rho_i)_\zeta(t)),
\end{equation*}
where $\zeta^{\bbD_i^\ell}$ is the matrix corresponding to the endomorphism $\zeta^{D_i^\ell}$. Explicitly, we have
\begin{gather*}
\zeta^{\bbD_1^\ell} = q^{\Phi_1 s/2} \left( \begin{array}{cccc}
\zeta^{- \ell s/4} & \\
& \zeta^{- (\ell - 2) s/4} \\
& & \ddots \\
& & & \zeta^{\ell s/4}
\end{array} \right), \\
\zeta^{\bbD_2^\ell} = q^{\Phi_2 s/2} \left( \begin{array}{cccc}
\zeta^{\ell s/4} & \\
& \zeta^{(\ell - 2) s/4} \\
& & \ddots \\
& & & \zeta^{- \ell s/4}
\end{array} \right).
\end{gather*}
It is worth to note here that
\begin{equation*}
(\rho_1)_\zeta(t) = q^{- (\Phi_1 - \Phi_2) N}, \qquad (\rho_2)_\zeta(t) = q^{- (\Phi_2 - \Phi_1) N}.
\end{equation*}

\subsection{Functional relations}

The functional relations for the transfer operators $\bbT^{k, \, \ell}(\zeta | \eta_1, \ldots, \eta_N)$ and $Q$-operators $\bbQ^\ell_k(\zeta | \eta_1, \ldots, \eta_N)$ can be obtained by acting on the universal functional relations by the corresponding representations of $\uqlslii$. Therefore, they have one and the same form. Note that the transfer operators and $Q$-operators usually contain some transcendental functions on the spectral parameter $\zeta$. From the point of view of concrete applications to integrable systems it is desirable to work with rational functions on $\zeta$, or on some fixed function on $\zeta$. A simple analysis shows that the $Q$-operators $\hbbQ^\ell_k(\zeta | \eta_1, \ldots, \eta_N)$ related to the $Q$-operators $\bbQ^\ell_k(\zeta | \eta_1, \ldots, \eta_N)$ by the equation
\begin{equation*}
\bbQ^\ell_k(\zeta | \eta_1, \ldots, \eta_N) = \zeta^{\Phi_k s/2} \left[ \prod_{i = 1}^N \prod_{j = 1}^\ell \zeta^{s/4} \exp \big( f_2(q^{2 (\ell - j)} \zeta^s \eta_i^{-s}) \big) \right] \hbbQ^\ell_k(\zeta | \eta_1, \ldots, \eta_N)
\end{equation*}
are Laurent polynomials on $\zeta^{s/2}$. Similarly, one can demonstrate that the transfer operators $\hbbT^{k, \, \ell}(\zeta | \eta_1, \ldots, \eta_N)$ related to the transfer operators $\bbT^{k, \, \ell}(\zeta | \eta_1, \ldots, \eta_N)$ by the equation
\begin{multline*}
\bbT^{k, \, \ell}(\zeta | \eta_1, \ldots, \eta_N) \\=\left[ \prod_{i = 1}^N \prod_{j = 1}^\ell q^{- k/2} \zeta^{s/2} \eta_i^{-s/2} \exp \big( f_2(q^{2 (\ell - j) - 2 k - 1} \zeta^s \eta_i^{-s}) + f_2(q^{2 (\ell - j) + 1} \zeta^s \eta_i^{-s}) \big) \right] \\ \times \hbbT^{k, \, \ell}(\zeta | \eta_1, \ldots, \eta_N)
\end{multline*}
are Laurent polynomials on $\zeta^{s/2}$ as well. In fact, in this case we have Laurent polynomials on $\zeta^s$.

In terms of the polynomial objects we have the following analogue of the $TQ$-relation from (\ref{tq})
\begin{multline*}
\hbbT^{1, \, \ell}(q^{1/s} \zeta | \eta_1, \ldots, \eta_N) \hbbQ_k^\ell(\zeta | \eta_1, \ldots, \eta_N) \\
= q^{\Phi_k} \left[ \prod_{i = 1}^N \prod_{j = 1}^\ell q^{\ell - j} \, b(q^{-(\ell - j) + 1/2} \zeta^{s/2} \eta_i^{-s/2}) \right] \hbbQ_k^\ell(q^{2/s} \zeta | \eta_1, \ldots, \eta_N) \\
+ q^{-\Phi_k} \left[ \prod_{i = 1}^N \prod_{j = 1}^\ell q^{\ell - j} \, b(q^{-(\ell - j) - 1/2} \zeta^{s/2} \eta_i^{-s/2}) \right] \hbbQ_k^\ell(q^{-2/s} \zeta | \eta_1, \ldots, \eta_N).
\end{multline*}
Here and below we use the notation
\begin{equation*}
b(\zeta) = \zeta - \zeta^{-1}.
\end{equation*}
Relations (\ref{fr1}) and (\ref{fr2}) lead to the $TT$-relation
\begin{multline*}
\hbbT^{k, \, \ell}(q^{-1/s} \zeta | \eta_1, \ldots, \eta_N) \hbbT^{k, \, \ell}(q^{1/s} \zeta | \eta_1, \ldots, \eta_N) \\
= \prod_{i = 1}^N \prod_{j = 1}^\ell q^{2(l - j)} \, b(q^{- (\ell - j) + k + 1/2} \zeta^{-s/2} \eta_i^{s/2})  \, b(q^{- (\ell - j) - 1/2} \zeta^{-s/2} \eta_i^{s/2}) \\
+ \hbbT^{k - 1, \, \ell}(q^{-1/s} \zeta | \eta_1, \ldots, \eta_N) \hbbT^{k + 1, \, \ell}(q^{1/s} \zeta | \eta_1, \ldots, \eta_N)
\end{multline*}
and to the $TT$-relation
\begin{multline*}
\hbbT^{1, \, \ell}(q^{-2k/s} \zeta | \eta_1, \ldots, \eta_N) \hbbT^{k, \, \ell}(\zeta | \eta_1, \ldots, \eta_N) \\*
= \left[ \prod_{i = 1}^N \prod_{j = 1}^\ell q^{\ell - j} \, b(q^{-(\ell - j) + k + 1} \zeta^{s/2} \eta_i^{-s/2}) \right] \hbbT^{k - 1, \, \ell}(\zeta | \eta_1, \ldots, \eta_N) \\
+ \left[ \prod_{i = 1}^N \prod_{j = 1}^\ell q^{\ell - j} \, b(q^{-(\ell - j) + k} \zeta^{s/2} \eta_i^{-s/2}) \right] \hbbT^{k + 1, \, \ell}(\zeta | \eta_1, \ldots, \eta_N).
\end{multline*}

\section{Conclusions}

We have rederived the universal functional relations for the quantum integrable systems related to the quantum group $\uqlslii$ treating the Jimbo's homomorphism as a mapping from $\uqlslii$ to $\uqglii$. Comparing with the case when the Jimbo's homomorphism is considered as a mapping from $\uqlslii$ to $\uqslii$ we see that the arising formulas are simpler. The obtained relations together with the results obtained for the systems related to the quantum group $\uqlsliii$ \cite{BazHibKho02, BooGoeKluNirRaz14b} make a generalization of the functional relations to the case of the quantum group $\mathrm U_q(\calL(\mathfrak{sl}_n))$ almost evident. We have derived expressions for monodromy operators and $L$-operators for the case of `spins' $1/2$ and $1$. Finally, we have rewritten the functional relations for the case of chains of arbitrary `spin' particles in terms of polynomial objects. This question was also treated earlier in the papers \cite{KirRes87, Man14} by different approaches.

\vskip .5em

{\em Acknowledgements.\/} This work was supported in part by the DFG grant KL \hbox{645/10-1}, the RFBR grants \#~13-01-00217, \#~14-01-91335, and by the Volkswagen Foundation. We thanks our colleagues and coauthors H. Boos, F. G\"ohmann and A. Kl\"umper for numerous discussions. A. V. R. would like to thank the Wuppertal University for hospitality.

\newcommand{\noopsort}[1]{}
\providecommand{\bysame}{\leavevmode\hbox to3em{\hrulefill}\thinspace}
\providecommand{\href}[2]{#2}
\providecommand{\curlanguage}[1]{%
 \expandafter\ifx\csname #1\endcsname\relax
 \else\csname #1\endcsname\fi}

\end{document}